\documentclass[aps,pra,showpacs,twoside,twocolumn,longbibliography,10pt]{revtex4-1}
\usepackage[colorlinks=true, citecolor=red, urlcolor=blue ]{hyperref}
\usepackage{epsfig,newlfont,amssymb,amsfonts,amsmath,bm,subfigure,palatino,mathtools,amsthm,braket,times,soul,enumitem,color}
\usepackage[normalem]{ulem}
\newcommand{\stkout}[1]{\ifmmode\text{\sout{\ensuremath{#1}}}\else\sout{#1}\fi}
\usepackage[english]{babel}
\usepackage[utf8]{inputenc}
\usepackage{array}
\usepackage{graphics}

\def\Tr{\text{Tr}}

\begin{document}

\title{Nonlinearity-assisted advantage for open Dicke-quantum batteries}


\author{Aparajita Bhattacharyya, Pratha Dongre, Ujjwal Sen}
\affiliation{Harish-Chandra Research Institute, A CI of Homi Bhabha National Institute, Chhatnag Road, Jhunsi, Prayagraj 211 019, India}

\begin{abstract}
We analyze the performance of a quantum battery in terms of energy storage and energy extraction, assisted by nonlinearities in a Dicke quantum battery utilizing an open-system approach. In particular, we consider two types of nonlinearities in the system, viz. nonlinearity in the coupling between battery and the cavity, and the cavity itself comprising an anharmonic oscillator. In both these scenarios, the cavity is connected to an environment, and is driven by an external laser source. We derive the Markovian master equation for the dynamics of the combined cavity-battery system in presence of the environment.  When the cavity is non-linearly coupled to a battery, we find an enhancement in the steady state ergotropy over the linearly-coupled case. We further see that the times at which steady state  ergotopy is attained get decreased in the presence of nonlinear coupling. We also identify instances where no ergotropy is obtained in the linear case, but can be obtained in presence of any non-zero value of nonlinearity. We additionally find a complementarity between the nonlinear interaction strength and the coherent-drive strength in the steady-state ergotropy values, i.e. the same ergotropy as that obtained using weak nonlinearity and strong coherent drive can be obtained utilizing an opposite order of their values.
For completeness, we also identify the nonlinear interaction where the cavity can be considered as a ``charger" and demonstrate an advantage of using the nonlinear coupling over linear ones.
In  case when the cavity is modeled as a multi-level transmon with an inherent anharmonicity, we find an advantage in maximum ergotropy values over the harmonic instance. Unlike the non-linearly coupled batteries, the ones mediated by an anharmonic cavity prove to be useful in the transient regime. We also determine values of the coherent drive for which the anharmonic battery is beneficial over the harmonic one at all timescales. Further, we observe a distinct region of the anharmonic strength and coherent drive where ergotropy values reach their maxima.

\end{abstract}

\maketitle

\section{Introduction}
In the last two decades, there have been significant developments in quantum technologies, for instance, in transistors~\cite{bardeen1949physical,pearson1955history}, lasers~\cite{bromberg1988birth} etc. which in the recent years have facilitated the study of individual quantum states and allowed us to manipulate systems exhibiting quantum properties such as superposition and entanglement.
The major avenues of quantum technology that are currently achieving immense practical importance, are
: quantum computation~\cite{Preskill2018quantumcomputingin}, quantum communication~\cite{gisin2007quantum}, quantum simulation~\cite{blatt2012quantum} etc.
This push towards the development of quantum technologies can be viewed from the perspective of two driving forces. Firstly, we identify that the increasing need to miniaturize technology has elicited the consideration of nontrivial effects of quantum mechanics. This emerging property calls for a new understanding of concepts in traditional thermodynamics, such as work, heat, and entropy ~\cite{alicki1979quantum,deffner2019quantum,Millen_2016}. The second motivation is the potential advantage achieved by harnessing quantum effects in certain applications, such as the enhancement of certain thermodynamic tasks in quantum machines, such as heat engines ~\cite{PhysRevLett.2.262,kosloff1984quantum, saha2023}, refrigerators~\cite{popescu,PhysRevLett.123.170605,mitchison2019,apa_ref}, and the quantum batteries ~\cite{Acin,Frey,PhysRevLett.118.150601,PhysRevLett.120.117702,PhysRevB.98.205423,Alhambra,Santos,Sun,Srijon1,Polini0,Srijon2,Kornikar,Tanoy2,fischer}. 

The storage of energy and subsequently extracting parts of the stored energy has become an important subject of study in quantum thermodynamics due to its practical importance ~\cite{allahverdyan2004maximal,perarnau2015extractable,binder2015quantacell}. The study of quantum batteries - quantum mechanical systems designed to store energy and extract it by harnessing quantum effects - has extensively been conducted in the recent years ~\cite{PhysRevLett.111.240401,le2018spin,andolina2019extractable,ncp_Aparajita}. The extraction of this stored energy from a quantum battery by employing unitary processes has been extensively employed. The maximum amount of energy that can be withdrawn from a quantum battery using unitary operations is referred to as the ergotropy of the system~\cite{alicki2013entanglement}. The state of the system from which no further energy can be squeezed out, i.e., the states that cannot be further discharged  under unitary evolution, are called “passive” states. Such states are known to be diagonal in the energy eigenbasis of the Hamiltonian and ordered with decreasing eigenvalues corresponding to increasing energies.

Sophisticated theoretical proposals have been put forth that aim to increase the efficiency and performance of quantum batteries by exploiting inherent quantum advantage ~\cite{crescente2020charging,PhysRevE.101.062114,carrega2020dissipative,PhysRevE.103.042118,shaghaghi2022micromasers,PhysRevLett.128.140501,chaki2023,chaki2024,cata}.
Previous analyses have established the advantage of coherent charging in a hybrid system, i.e. a harmonic oscillator as auxiliary and two-level system as battery, and have concluded that the time at which energy and ergotropy remain maximal decreases monotonically with the external driving field~\cite{andolina2019charger}. 
It has also been explored how a coherent drive affects the charging of such a hybrid battery,
identifying that the coherent drive aids in increasing energy extraction efficiency~\cite{downing2023quantum}.

The study of light-matter interaction in the context of quantum batteries has gained utmost attention, particularly within platforms like circuit quantum electrodynamics (circuit-QED), where interactions between solid-state qubits and quantized cavity fields can be precisely engineered and controlled. 
Such scenarios are physically realizable in several physical systems such as in superconducting circuits~\cite{Felicetti_2018}, trapped ion systems~\cite{Felicetti_2015}, etc. 
In such systems, the Dicke model serves as a foundational framework, capturing the collective dynamics of two-level systems coupled to a single-mode cavity field. The system under investigation here is inspired by this paradigm, but tailored to the minimal yet powerful case of a single qubit interacting with a cavity through both single-photon~\cite{PhysRevB.98.205423,PhysRevLett.120.117702} and two-photon~\cite{PhysRevA.95.053854,Garbe2020} coupling channels. A coherent laser drive introduces an external energy source, modulating the cavity dynamics and allowing for active control of the charging process. 
Additionally, the cavity is subject to the influence of a Markovian environment.
In order to belong to the Markovian
family of environments, the thermal environments must be infinitely large and
have a continuous energy spectrum. The bosonic environment, consisting of an infinite number of harmonic oscillators, within certain constraints, behaves as a Markovian one.

The fundamental motivation of this work is to investigate how nonlinear light-matter interactions, particularly two-photon coupling, can enhance the performance of qubit batteries compared to the conventional linear interactions, i.e. Jaynes-Cummings type.
In the first part our results, the cavity is modeled as a harmonic oscillator with energy spacing set by a reference frequency, and the gap between the two levels of the qubit is tuned to the same frequency to allow for resonant energy exchange. 
The total Hamiltonian includes contributions from the local energies of the cavity and qubit, the coherent driving of the cavity, and both linear and nonlinear interaction terms. 
By moving to the interaction picture and applying the rotating-wave approximation, the resulting interaction Hamiltonian depicts the interplay between resonant single-photon and two-photon processes. This setup is particularly relevant for quantum battery models, where understanding how energy can be stored and transferred efficiently via engineered nonlinear interactions is of both fundamental and practical interest.
 There can arise two scenarios - first, when the cavity is accessible and energy can be extracted directly from the cavity, and secondly, when one does not have access to the cavity and energy is to be extracted from the battery. We mainly focus on the second case, where the cavity is not accessible for energy extraction. Our objective is to investigate how the presence of both linear and nonlinear interactions between the battery and the auxiliary influences the energy extraction process from the battery. 
We identify instances where the nonlinearity assisted charging enhances the steady-state ergotropy, and facilitates faster achievement of the relevant steady-state. We also identify instances where no ergotropy is obtained in the linear case, but can be obtained in presence of any non-zero value of nonlinearity. We also observe a complementary nature of the steady-state ergotropy values with respect to the nonlinear interaction and the coherent drive strengths. In this part of our analysis, we show that the Dicke quantum battery model with single- and two-photon couplings can facilitate energy extraction in the steady state, i.e. at relatively large timescales when the system attains equilibrium.
Previously, the effects of nonlinear interactions such as in two-photon process in a quantum battery has been considered, as given in in~\cite{delmonte2021characterization,PhysRevB.102.245407}, although the main focus in these works was on closed system dynamics, i.e. in the absence of an environment.

The second part of our results comprises the scenario when the interaction between battery and cavity is linear, however the cavity is considered to an anharmonic oscillator, particularly a transmon. We make the assumptions of Markovian evolution~\cite{BRE02,rivasarxivcitation} for the dynamics of the combined system of the battery and the anharmonic auxiliary, when the auxiliary is locally connected to an environment.
 In the anharmonicity-assisted charging, we find enhancement of the maximum ergotropy values, and find regimes of the coherent drive strength when the ergotropy is higher in the anharmonic case than that in the harmonic one at all timescales.  
Anharmonic oscillator systems can be experimentally implemented using transmons. The system is characterized by a set of canonically conjugate quantum operators derived from the count of Cooper pairs moved through the junction and the phase difference across it. Charge-based systems are susceptible to stray electric field disturbances. This undesirable condition can be significantly mitigated by operating the Cooper pair box in the ``transmon" regime, where the Josephson tunneling energy prevails over the Coulomb charging energy~\cite{PhysRevLett.89.117901,PhysRevLett.95.210503,wendin2017quantum}.

The remainder of the paper is arranged as follows. The relevant information necessary to formulate the problem is discussed in Sec.~\ref{prelim}. This includes a discussion on the quantum battery model that we consider, and extraction of energy from quantum batteries. In Sec.~\ref{sec:Non-lin_coupling}, the master equation arising as a result of the interaction of the auxiliary with the environment, and the performance of quantum battery in energy extraction are provided, if the battery is non-linearly coupled to the auxiliary. The dynamics of the auxiliary-battery system, and its effects in energy extraction, in presence of anharmonic auxiliary linearly coupled to a battery, is provided in Sec.~\ref{sec:anharmonic}. Finally, the concluding remarks are presented in Sec.~\ref{concl}.

\section{Preliminaries}
\label{prelim}
In this section, we describe the Dicke model which consists of a battery coupled to a  cavity mode. A coherent laser source injects energy into the cavity, while the cavity is also kept in contact with an environment. Our aim is to extract as much energy as possible from the battery, utilizing the correlations between the cavity and the battery. Then, we discuss the prerequisites for the extraction of energy from quantum batteries using unitary operations.

\subsection{Quantum battery model}
The nonlinear Dicke quantum battery model comprises four distinct elements. The first element is a quantum battery, $B$, which is considered to be a qubit. The second element, $A$, is the cavity, also referred to as the auxiliary. The third element, $L$, is the external energy supply of the model, which can be visualized as a classical laser source that injects energy into the battery through the auxiliary. In addition, the auxiliary, $A$, is kept in contact with a thermal reservoir, $E$, which is considered to be bosonic and Markovian in nature. 
The auxiliary and battery are described by their corresponding local Hamilotonians, $H_A$ and $H_B$. They also interact with each other through an interaction $H_{AB}^{(1)}$.  The coherent laser source, $L$, is applied to the auxiliary at time $t=0$, and the local modulating Hamiltonian of $A$ due to $L$  is given by  $\Delta H_A(t)$. The Hamiltonian, $H_{EA}$, describes the interaction between the bosonic environment, $E$ and the auxiliary, $A$. 
So the global Hamiltonian of the  composite system is given by
\begin{equation}
    H = H_0 + \lambda(t) H_1,
\end{equation}
where $H_0 = H_A+H_B$ and $H_1=H_{AB}^{(1)}+H_{EA}+\Delta H_A(t)$. The  dimensionless quantity, $\lambda(t)$, takes value 1 when $t \in [0,\tau]$, and otherwise takes value zero. Physically it represents a classical parameter that acts as a switch and can be externally controlled to turn the interactions on and off. The two interaction terms, $H_{AB}^{(1)} $ and $ H_{EA}$, and the coherent drive, $\Delta H_A(t)$, are effective within the time period, $[0,\tau]$, after which they are all disconnected. 

We initially consider the cavity to be a harmonic oscillator with the energy difference between two consecutive levels being $\omega_0$. The battery is considered to be a single qubit, whose energy gap is exactly equal to the frequency of $A$. The frequency of the laser source is also considered to be $\omega_0$.
The local Hamiltonians of $A$ and $B$, and the modulating Hamitonian of $A$ due to $L$ are given by 
\begin{equation}
    \label{nonLinHamiltonian_HO_qubit}
    \begin{split}
        H_A &= k_2 \omega_0 a^{\dagger} a,\\
    H_B &=K \frac{\omega_0}{2} (\sigma^z_B+1),\\
    \Delta H_A(t) &=k_1 F(e^{-i\omega_0 t}a^{\dagger}+e^{i\omega_0 t}a),
    \end{split}
\end{equation}
where $k_1=K\sqrt{\tilde{\omega}}$ and $k_2=K\tilde{\omega}$ represent the dimensions.
Here, $a$($a^{\dagger}$) are the bosonic annihilation (creation) operators of the system, $A$, having the unit of $1/\sqrt{\tilde{\omega}}$. The quantity, $\tilde{\omega}$, has magnitude one, and has the dimension of frequency. This notation is followed throughout the paper to denote the unit of frequency.  The parameter, $K$,  has the dimension of energy and $\omega_0$ is dimensionless. The dimensionless quantity, $F$ represents the field strength corresponding to the coherent drive acting on $A$. Here  $\tilde{\omega}$ represents an arbitrary constant with the dimensions of frequency.

\subsection{Energy extraction using unitaries}
A quantum battery is a quantum mechanical system characterized by a density matrix and a Hamiltonian. The system can be charged by applying some unitary or non-unitary operations on it. Once it is charged, our aim is to extract maximum possible energy from the battery. The maximum extractable energy from a quantum battery, under unitary operations, is referred to as the ergotropy of the battery.
After the extraction of maximum amount of energy from the battery, the states from which no further energy can be extracted are termed as passive states.
In other words, passive states are states from which energy extraction is not possible using unitary operations. Passive states are important in the framework of quantum batteries or, more generally, in quantum thermodynamics~\cite{Silva,Huber,Brown_2016,Sparaciari_2017,Kalu}.

In the case that we consider, the auxiliary interacts with the coherent laser source and also with the environment, as a result of which average energy of the auxiliary varies in the time interval, $[0,\tau]$. Further, due to the interaction of the auxiliary with the battery, exchange of energy occurs between $A$ and $B$ as well. Therefore, as the laser injects energy into the auxiliary, the battery also gets charged due to the presence of the interaction, $H_{AB}$, in the time interval $[0,\tau]$. The average energy contained in $B$ at the end of the charging process is $E_B(\tau) = tr[H_B\rho_B(\tau)]$. Our aim is to extract the maximum energy from $\rho_B(\tau)$ using unitary operations.
The ergotropy, $\varepsilon_B(\tau)$, in such a case, is mathematically given by
\begin{eqnarray}
\label{erg}
   \varepsilon_B(\tau) &=& \text{Tr}(\rho_B(\tau) H_B) - \min_U \text{Tr}\left[U(\lambda)\rho_B(\tau) U(\lambda)^{\dagger} H_B\right], \nonumber \\
    &=& \text{Tr}(\rho_B(\tau) H_B) -  \text{Tr}(\rho^{(p)}_B H_B)\nonumber \\
    &=& E_B(\tau) - E_B^{(p)} (\tau),
\end{eqnarray}
where the first term on the right-hand side of the last equation represents the average energy of $\rho_B$ at time $\tau$, and the quantity, $ E_B^{(p)} (\tau) = tr[H_B \rho^{(p)}_B(\tau)],$ is the energy of the passive state, $ \rho^{(p)}_B$ corresponding to state $ \rho_B$.
So the amount of energy contained in the passive state,  $\rho^{(p)}_B$, is  ``forbidden", in the sense that, this energy cannot be extracted from $\rho_B(\tau)$, for the given $H_B$ and would be left in the system after extracting all the unitarily accessible energy from $\rho_B(\tau)$.
The necessary and sufficient conditions for a state, $\rho^{(p)}_B$, to be passive are that it should commute with its Hamiltonian, $H_B$, and if for a particular order of the eigenvectors the corresponding eigenvalues of $H_B$, $\{\epsilon_i\}_i$, satisfy $\epsilon_i>\epsilon_j$, then the eigenvalues of the passive state, $\{\lambda\}_i$, should satisfy $\lambda_i\leq \lambda_j$ for all $i$ and $j$~\cite{Lenard,Pusz}.

In each of the above cases, we consider two scenarios, i.e. one where the cavity is accessible and energy can be extracted from the cavity, and the second where the cavity is not accessible and only the battery is available for energy extraction. 

\section{Open quantum battery nonlinearly coupled to cavity}
\label{sec:Non-lin_coupling}

The system under consideration is described by the well-known Dicke model. Specifically, we study a two-level system coupled to a single-mode cavity field through both single-photon and two-photon interaction processes. Additionally, the cavity is driven by an external coherent laser field. The total Hamiltonian of the system is given by
\begin{equation}
    H = H_A + H_B + \Delta H_A(t) + \widetilde{H}_{AB}^{(1)} + \widetilde{H}_{AB}^{(2)},
\end{equation}
where \( \widetilde{H}_{AB}^{(1)} \) and \( \widetilde{H}_{AB}^{(2)} \) represent the single-photon and two-photon interaction terms between the qubit and the cavity mode, respectively. The other terms in the Hamiltonian are described previously. The interaction terms are expressed as
\begin{align}
    \widetilde{H}_{AB}^{(1)} &= K\sqrt{\tilde{\omega}}\, g_1 (a + a^{\dagger})(\sigma^{+} + \sigma^{-}), \\
    \widetilde{H}_{AB}^{(2)} &= K\tilde{\omega}\, g_2 (a + a^{\dagger})^2 (\sigma^{+} + \sigma^{-}),
\end{align}
where \( g_1 \) and \( g_2 \) denote the dimensionless coupling strengths corresponding to the linear (single-photon) and nonlinear (two-photon) interactions, respectively, and \( K \) and \( \tilde{\omega} \) are defined previously.
Moving to the interaction picture, and applying the rotating wave approximation, the final interaction Hamiltonian is given by 
\begin{eqnarray}
    H_{AB}= k_1 g_1 (a\sigma^{+} + a^{\dagger}\sigma^{-})+k_2 g_2 (a^2\sigma^{+} + {a^{\dagger}}^2\sigma^{-}), \;\;
    \label{interaction}
\end{eqnarray}
where $k_1=K\sqrt{\tilde{\omega}}$ and $k_2=K\tilde{\omega}$ take care of the dimensions.
%
The linearly-coupled Hamiltonian, $H_{AB}^{(1)}=g_1 (a\sigma^{+} + a^{\dagger}\sigma^{-})$ commutes with the sum of local energies of the battery and auxiliary. So the linear interaction does not change the total local energies of the battery and auxiliary. We examine the situation of introducing nonlinearities to the system.

\subsection{Interaction with the environment}
\label{def}
The interaction between the auxiliary system, $A$, and the bosonic Markovian environment, $E$, is described by the Hamiltonian,
\begin{equation}
    \label{HO_bath_hamiltonian}
    \begin{split}
        H_{EA} &= \int_0^{\omega_{max}} \sqrt{\tilde{\omega}} \hbar d\omega h(\omega) \left(a_{\omega}a^{\dagger}+ a^{\dagger}_{\omega}a \right),
    \end{split}
\end{equation}
where the operators $a^{\dagger}_{\omega}(a_{\omega})$, having the unit of $1/\sqrt{\tilde{\omega}}$, represents the bosonic creation (annihilation) operators corresponding to the mode $\omega$ of the bath.
Here $\omega_{max}$ denotes the cutoff frequency of the bath. This cutoff frequency is set to be sufficiently high so that the memory time of the bath $\sim \omega_{max}^{-1}$, is negligibly small.
For the numerical analyses, the cutoff frequency, $\omega_{max}$, is considered to be equal to $1000$, in units of $\tilde{\omega}$, uniformly throughout the paper.
This choice of parameters allows us to apply the Markovian approximations~\cite{BRE02, AlickiLendi2007, rivasarxivcitation,lidar2020lecture}. Here $h(\omega)$ is a dimensionless function of $\omega$, that tunes the coupling between the harmonic oscillator and environment. The local Hamiltonian describing the bath is given by
\begin{eqnarray}
H_E=\int_0^{\omega_{max}} \hbar \tilde{\omega} a_{\omega} a_{\omega}^{\dagger} d\omega.
\end{eqnarray}
For harmonic oscillator baths, $\tilde{\omega} h^{2}(\omega)=J(\omega)$, where $J(\omega)$ represents the spectral density function of the bath. In this paper, we have taken $J(\omega)$ to be the Ohmic spectral density function having the form $J(\omega)= \alpha \omega \exp(-\omega/\omega_{max})$. Here $\alpha$ represents the dimensionless interaction strength between the harmonic-oscillator auxiliary and the bosonic bath.

Each of $A$ and $B$ are assumed to be initially prepared in the ground states of their respective Hamiltonians, denoted by $|0 \rangle \langle 0|_A $ and $ |0 \rangle \langle 0|_B$,  such that the joint auxiliary-battery initial state is $\rho_{AB}(t=0) = |0 \rangle \langle 0|_A \otimes |0 \rangle \langle 0|_B$.
The quantity, $|0 \rangle \langle 0|_B$ is simply the ground state of the Hamiltonian, $H_B$, whereas $|0 \rangle \langle 0|_A$ represents the ground state of the operator, $\omega_0 a^{\dagger} a$.
As we turn on the dissipation between $A$ and $E$, the coherent drive, $\Delta H_{A}$, and the interaction between $A$ and $B$, the joint system of $EAB$ evolves unitarily under the global Hamiltonian, $H$, following which, the environment is discarded. Since the initial system-environment state - the system comprising  the auxiliary and battery - is product, and a global unitary acts on the entire system-environment state, followed by tracing out the environment, the evolution of the system can be effectively considered to be a completely positive trace preserving (CPTP) operation. Moreover, Markovian approximations are made while considering such a CPTP operation on the  auxiliary and battery. Therefore the dynamics of the combined state of the battery and auxiliary is governed by the the Gorini-Kossakolski-Sudarshan-Lindblad (GKSL) master equation.
Throughout the process, energy is transferred between $A$ and $E$, and due to the interaction between $A$ and $B$, energy is also exchanged between $A$ and $B$ as well. As a result, the dynamics of $B$, considered separately, becomes non-Markovian in nature.

To simplify the analysis of the system dynamics, we move to the interaction picture representation. The resultant density matrix of $A$ and $B$ at time $\tilde{t}$, given by $\rho_{AB}(t)$ in the interaction picture, is
 \begin{equation}
     \rho_{AB}(t) = e^{i(H_A+H_B)\tilde{t}/\hbar}\rho_{AB}(\tilde{t}=0)e^{-i(H_A+H_B)\tilde{t}/\hbar}.
 \end{equation}
 Here, $t$ denotes the dimensionless time with the actual time $\tilde{t}$, defined as $t=\frac{K \tilde{t}}{\hbar}$. 
 In the interaction picture, the evolution of such a system within the time period, $t\in [0,\tau]$, is governed by the following master equation
 \begin{equation}
 \label{lindblad}
    \Dot{\Tilde{\rho}}_{AB}(t) = \mathcal{L}_{AB}[\rho_{AB}(t)],
\end{equation}
 where $ \mathcal{L}_{AB}$ is the GKSL super-operator ~\cite{Kossakowski1972,10.1063/1.522979,Lindblad1976,lidar2020lecture,Manzano_2020}. 
 The term, $\mathcal{L}_{AB}[...]$, can explicitly be written as
 \begin{eqnarray}
 \label{dissipator}
     \mathcal{L}_{AB}(t)[\rho_{AB}(t)] = &-&i \frac{1}{K} [\Delta H_A (t)+ H_{AB}, \rho_{AB}(t)]\nonumber \\
     &+& \frac{\hbar}{K}  \mathcal{D}\left[\rho_{AB}(t)\right],
 \end{eqnarray}
 where the first term on the right hand side of Eq.~\eqref{dissipator} represents the unitary part, whereas the second term is the non-unitary dissipator term. 
The dynamical equation for the joint system $AB$, is described by the GKSL master equation, as presented in Eq.~\eqref{lindblad}, with the dissipative term given by
\begin{eqnarray}
\label{lindblad2}
\mathcal{D}\left[\rho_{AB}(t)\right] &=&  \sum_{\omega^{\prime}} \gamma(\omega^{\prime}) \Big[L^{\omega^{\prime}} \rho_{AB}(t) L^{\omega^{\prime\dagger}} \nonumber \\ 
    &-& \frac{1}{2} \{ L^{\omega^{\prime\dagger}} L^{\omega^{\prime}},\rho_{AB}(t) \} \Big].
\end{eqnarray}
Here the quantity, $\gamma(\omega^{\prime})$ denotes the dimensionless decay constant, where $\tilde{\gamma}(\omega^{\prime})=\frac{\hbar \gamma(\omega^{\prime})}{K}$. The operators, ${L^{\omega^{\prime}}_i }$ represent the Lindblad or jump operators associated with the possible transition frequencies $\omega^{\prime}$ of the joint state of $A$ and $B$.
In the Markovian master equation, the information of the reservoirs is embedded in the transition rates or decay constants,  $\{ \tilde{\gamma}(\omega^{\prime}) \}$, which is given by
\begin{eqnarray}
\label{gamma}
\tilde{\gamma}(\omega^{\prime}) &=& J(\omega^{\prime}) [1+f(\omega^{\prime},\beta)] \quad \quad  \omega^{\prime}>0 \nonumber \\
                  &=& J(|\omega^{\prime}|) f(|\omega^{\prime}|,\beta) \quad \quad \quad   \omega^{\prime}<0,
\end{eqnarray}
where $f(\omega^{\prime},\beta)=[\exp(\beta \hbar \omega^{\prime})-1]^{-1}$ is the Bose-Einstein distribution function for the bosonic heat bath.
The value of the inverse temperature, $\beta$, chosen for numerical purposes throughout the paper is equal to $1$ in units of $k_B/K$, where $k_B$ is the Boltzmann constant and $K$ has the dimension of energy.
The joint evolution of the auxiliary and battery can be described by the Markovian master equation given in Eq.~\eqref{lindblad}, by solving which, one can obtain the density operator, $\rho_{AB}(t)$, as a function of time. Following this, the auxiliary is traced out to obtain the dynamics of the battery, $B$, alone. Therefore, the evolution of the battery, considered separately, is in general, not Markovian in nature.

For solving the differential equations, we employ numerical methods
using C++, adhering to the truncation of the annihilation and creation operators for studying
the evolution of the system for a finite duration of time. In this paper, we have truncated the oscillators up to 10 levels. While truncating the matrices, it has been ensured that the differences in the higher eigenvalues become relatively small, and the relevant physical quantities converge for the energy-level cut-off chosen.

\subsection{
Importance of an auxiliary system in environment-assisted energy extraction}
The presence of a cavity, i.e. an auxiliary system, coupled with the battery, rather than directly addressing the battery  with a laser field, is a key feature of the model and is motivated by the desire to explore potentially advantageous modes of battery performance. In order to demonstrate the benefit of using an auxiliary coupled to a  battery in the course of energy extraction from the battery, we consider a scenario in the absence of an auxiliary system, where the battery itself is kept in a bosonic Markovian environment and is also driven by a coherent source. 
The battery Hamiltonian is given by $H_B =K \frac{\omega_0}{2} (\sigma^z_B+1)$, where $K$ denotes the energy dimension. The Hamiltonians describing the driving field, and the interaction of the battery with the environment are respectively given by
\begin{eqnarray}
    \Delta H_B&=&K\widetilde{F}(e^{-i\omega_0 t}\sigma^{+}+e^{i\omega_0 t}\sigma^{-}), \;\; \text{and}\nonumber \\
    H_{BE}&=&\int_0^{w_{max}}\sqrt{\tilde{\omega}} \hbar d\omega h(\omega)(\sigma^+a_{\omega}+\sigma^-a_{\omega}^{\dagger}),
\end{eqnarray}
where $\widetilde{F}$ is a dimensionless quantity which denotes the strength of the coherent drive.
We evaluate the erogotropy, i.e. the maximum extractable energy from the battery in this scenario, and compare it with the ergotropy obtainable in the case where the battery is linearly or nonlinearly coupled to the auxiliary.
The steady-state ergotropy of the battery in the absence of an auxiliary is given by $\varepsilon_B^{0}=0.0212$, for the values of coherent driving strength, $F=0.5$, and linear interaction strength, $g_1=0.1$. For the same values of $F$ and $g_1$, the steady-state ergotropy of the battery in the linearly and nonlinearly coupled cases are given by $\varepsilon_B^{1}=0.0821$ and $\varepsilon_B^{2}=0.1969$. The nonlinear interaction is considered to be $H_{AB}$, which comprises both single- and the two-photon couplings. Therefore, $\varepsilon_B^{0}<\varepsilon_B^{1}<\varepsilon_B^{2}$, which suggests that the ergotropy of the battery without an auxiliary is significantly lower than in the scenario where the auxiliary is linearly or nonlinearly coupled to the battery. 
So, the auxiliary-mediated interaction opens up the possibility of enhancing the battery performance in terms of maximum extractable energy, which would not be otherwise accessible in a direct-drive scenario.
Furthermore, it allows us to separate the source of energy, which is classical and potentially noisy, from the target battery, thus reducing the direct influence of classical noise on the stored energy.  

\subsection{Advantage of two-photon coupling over one-photon coupling in energy extraction}
\label{non_invariant_energy}

\begin{figure*}
\includegraphics[width=5.9cm]{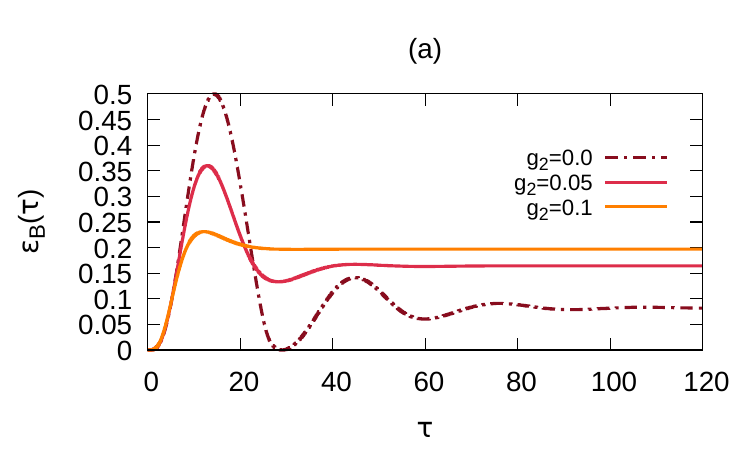}
\includegraphics[width=5.9cm]{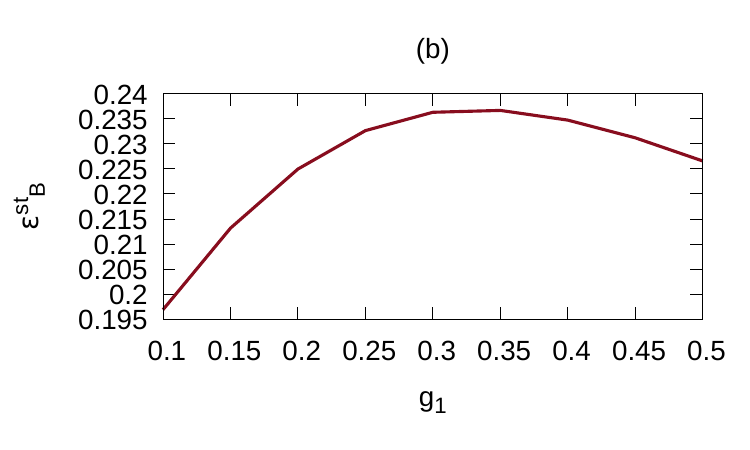}
\includegraphics[width=5.9cm]{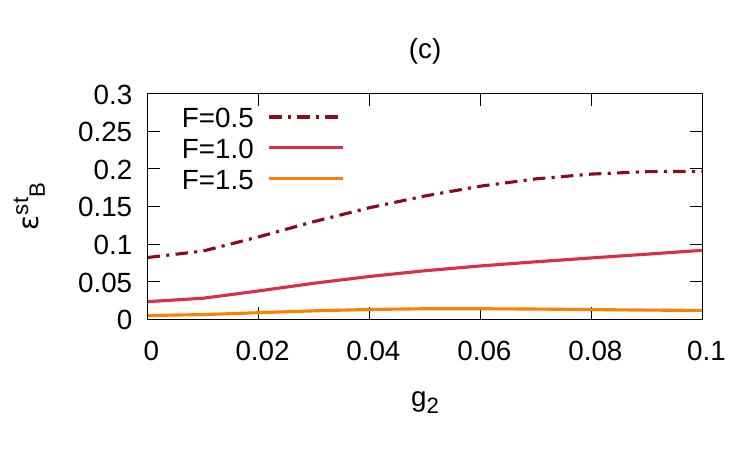}
\caption{\textbf{Panel (a):} Depiction of the maximum extractable energy versus time for different values of nonlinear interaction strength, $g_2$. The field strength corresponding to the coherent drive is $F = 0.5$, and the linear coupling strength is $g_1=0.1$.
Both in the main figure and inset, the quantities plotted along the horizontal axis is dimensionless, while ergotropy is units of $K$. 
\textbf{Panel (b):} Demonstration of the steady-state ergotropy as a function of the linear-interaction strength, $g_1$, for fixed value of coherent drive $F=0.5$ and nonlinear coupling $g_2=0.1$. The quantities plotted along both the axes are dimensionless.
\textbf{Panel (c):} The ergotropy of the battery, demonstrated as a function of the nonlinear interaction strength, $g_2$ for a different values of the driving field strength. The linear coupling strength is taken to be $g_1=0.1$. 
The quantities plotted along the horizontal axis is dimensionless, while ergotropy is in units of $K$.} 
\label{nonlinear_charg}
\end{figure*}

In this subsection, we investigate the maximum extractable energy from quantum batteries when the battery couples with an auxiliary using single- and two-photon couplings, as described by the Hamiltonian in Eq.~\eqref{interaction}. 
After obtaining the solution of the master equation for the joint density matrix \( \rho_{AB}(t) \),  the auxiliary is discarded and the ergotropy of the battery is computed using Eq.~\eqref{erg}.  
The resultant master equation in this case is of the following form
\begin{eqnarray}
    \label{nonLinHO_QB_ME}
        \dot{\Tilde{\rho}}_{AB}(t) = &-&i [\sqrt{\tilde{\omega}} g_1(a \sigma^{+} + a  \sigma^-) \nonumber \\
        &+&\tilde{\omega} g_2(a^2 \sigma^{+} + (a^{\dagger})^2 \sigma^-) \nonumber \\
        &+& \sqrt{\tilde{\omega}} F(a^{\dagger}+a),\; \rho_{AB}(t)]  \nonumber \\
        &+&  J(\omega_0) (1+f) \Big[a \rho_{AB}(t) a^{\dagger}- \frac{1}{2} \{ a^{\dagger} a,\rho_{AB}(t) \} \Big] \nonumber \\
        &+& J(\omega_0) f \Big[a^{\dagger} \rho_{AB}(t) a - \frac{1}{2} \{ a a^{\dagger},\rho_{AB}(t) \} \Big].
\end{eqnarray}
Here, \( J(\omega_0) = \alpha \omega_0 \exp(-\omega_0/\omega_{\text{max}}) \) denotes the Ohmic spectral density function, and \( f(\omega_0, \beta) = [\exp(\beta \hbar \omega_0) - 1]^{-1} \) represents the Bose--Einstein distribution function. The parameter \( \alpha \) is a dimensionless constant characterizing the strength of the interaction between the auxiliary and the bosonic bath. For numerical simulations, \( \alpha \) is set to unity.


\subsubsection{Energy extraction when auxiliary is accessible}
\label{aux_access}

Here we inspect the scenario in which the cavity, referred to as the auxiliary, is directly available for energy extraction.
As part of our analysis, we solve the master equation for the joint density matrix \( \rho_{AB}(t) \), after which we discard the battery subsystem and evaluate the ergotropy of the cavity using Eq.~\eqref{erg}.
We find that the ergotropy of the auxiliary is higher than that of the battery in each of the linear and nonlinear coupling scenarios, since the auxiliary is directly driven by the coherent source of energy. 
The difference between the steady-state ergotropy of the cavity and battery in the linear and nonlinear coupling scenarios are respectively given by $\Delta \varepsilon_{\text{lin}}=1.416$ and $\Delta \varepsilon_{\text{nonlin}}=1.276$.
However in this case, the steady-state ergotropy of the cavity in the nonlinearly coupled regime, given by $\varepsilon_A^{\text{st(nonlin)}}=1.498$, is smaller than its steady-state ergotropy in the linearly coupled situation, given by $\varepsilon_A^{\text{st(lin)}}=1.473$ by a an amount $0.025$. This shows that here the nonlinear interaction between the battery and auxiliary provides a small advantage over the linearly-coupled case, if the auxiliary is accessible for energy extraction. These results correspond to the values of driving strength, $F=0.5$, linear coupling strength $g_1=0.1$, and nonlinear strength $g_2=0.1$. 

\subsubsection{Energy extraction when auxiliary is not accessible}\label{aux_not_access}

After solving the master equation for $\rho_{AB}(t)$, we discard the auxiliary and  calculate the ergotropy of the battery using Eq.~\eqref{erg}. In the main plot of Fig.~\ref{nonlinear_charg}-(a), we plot the ergotropy of the battery along the vertical axis versus time along the horizontal axis.
Here we observe that the oscillatory behavior of $\varepsilon_B(\tau)$ is reduced with nonlinearity.
Moreover, we find that the nonlinearly coupled auxiliary provides an advantage in the steady state regime. The steady-state ergotropy is higher for non-zero nonlinear coupling than that of the linearly coupled scenario. Further, the attainment of steady state is faster in presence of nonlinear interaction between auxiliary and battery. In the transient regime, there is no advantage, i.e. the maximum ergotropy attained in the transient regime decreases in presence of nonlinear coupling.  So, we conclude that the nonlinear interaction between cavity and battery proves to be beneficial in the steady state regime, and therefore if in any setup, there is a need of energy extraction continuously at large timescales, then the nonlinearity assisted coupling between the battery and auxiliary is convenient.
The linear coupling strength is considered to be $g_1=0.1$. So we see that there is an advantage in the steady-state ergotropy over the linear case both in the regimes of weak and strong nonlinear coupling, i.e. $g_2$ is close to $0$ and $0.1$ respectively.
Furthermore, we find that there exists a time, $\tau=30$, when no energy can be extracted from the battery resulting in zero ergotropy in the linear coupling case, but the maximum extractable energy at that time corresponding to nonlinear coupling with any positive coupling strength is non-zero. 

Now, referring to Fig.~\ref{nonlinear_charg}-(b), we observe that, if the linear interaction strength increases, keeping the strength of nonlinear interaction constant, the steady-state ergotropy ofthe battery, denoted by $\varepsilon^{st}_B$, gradually increases, reaches a maximum, and then decreases further, although the variation in the value of $\varepsilon^{st}_B$ is within a range $\sim 0.05$. This analysis is performed for a fixed value of the coherent drive strength, $F=0.5$. This suggests that the ergotropy in the steady state has a small effect on linear interaction strength, in the sense that the variation of the steady-state ergotropy as a function of $g_1$ is within a sufficiently small range of $\sim 0.05$, for the chosen set of parameters. 

\begin{figure}[thpb]
    \centering 
    \includegraphics[width=1\linewidth]{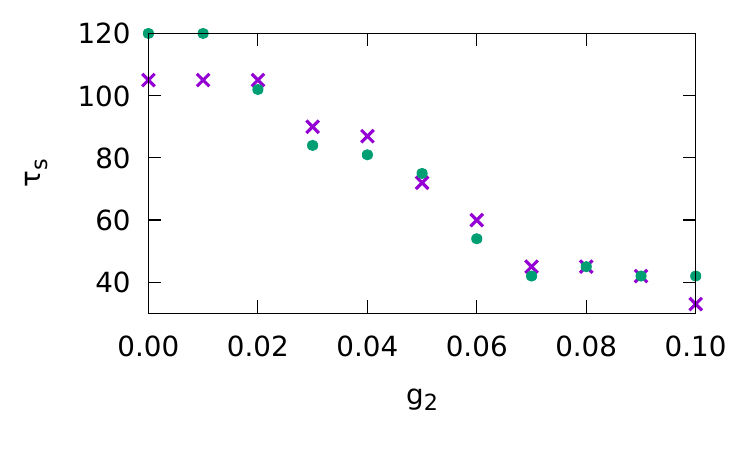} 
    \caption{The time $\tau_s$ at which the ergotropy,  $\varepsilon_B ( \tau )$,  and energy, $E_B(\tau)$, reaches their steady state values, as represented by teal circles and purple crosses respectively, which are plotted along the vertical axis,  versus the nonlinear interaction strength plotted along the horizontal axis. 
    Here we have obtained the results by setting $ g_1 = 0.1 $, $\gamma = 1.0$, $F =0.5$, and we have set inverse temperature $\beta = 1$. The quantities plotted along both the axes are dimensionless.} 
    \label{fig:nonlin_max_energy_ergo_diffg2_diffF} 
\end{figure}

Next let us look at the effect of varying the coherent drive strength on the steady-state ergotropy of the battery. In Fig.~\ref{nonlinear_charg}-(c), we demonstrate the steady-state ergotropy, $\varepsilon^{st}_B$,  as a function of the nonlinear interaction strength, $g_2$, for different values of the coherent drive, $F$. We find that the value of ergotropy in the steady state increases with an increase in the nonlinear strength, while the strength of coherent drive is within a range, $0.5 \le F \le 1.5$. The trend we find here is that the value of the steady-state ergotropy as well as the change in the steady-state ergotropy with $g_2$, decreases as we increase the coherent driving strength, $F$. So, it is intriguing to find that the maximum extractable energy in the steady state is enhanced with lower strength of coherent drive.

In Fig.~\ref{nonlinear_charg}-(c), we found that the the steady-state ergotropy increases with an increase in the nonlinear interaction strength between auxiliary and battery. Moreover, the time of attainment of steady state also becomes less as the nonlinearity is increased. In Fig.~\ref{fig:nonlin_max_energy_ergo_diffg2_diffF}, we plot the time where steady-state is reached, given by $\tau_s$, along the vertical axis versus the nonlinear strength, $g_2$, along the horizontal axis. We find that, with an increase in nonlinearity, the steady state is reached faster. So if there is a requirement of steady source of energy at comparatively small times, then the utility of nonlinearity in the interaction between auxiliary and battery proves to be useful.

\begin{figure}[thpb]
    \centering 
    \includegraphics[width=1\linewidth]{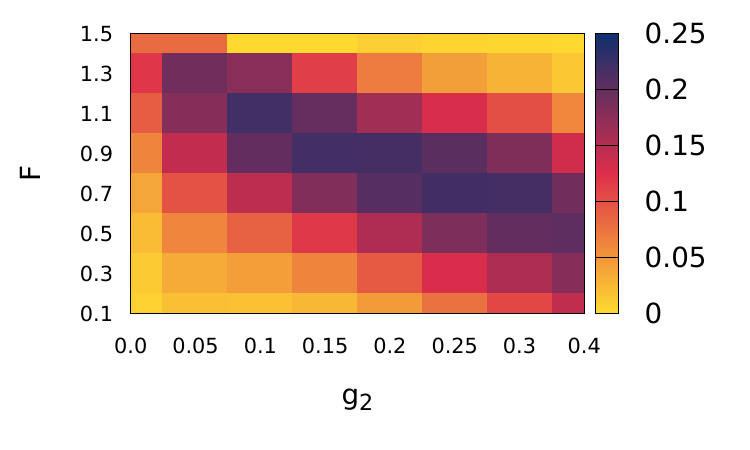} 
    \caption{Depiction of the steady state value of ergotropy obtained for different values of the coherent drive intensity $F$ along the vertical axis, and the nonlinear coupling strength, $g_2$, along the horizontal axis. The quantities plotted along the horizontal and vertical axes are dimensionless, while ergotropy is in units of $K$.} 
    \label{fig:new_max_g2_F_steady_heatmap} 
\end{figure}

Having seen that for lower values of the coherent drive, higher nonlinear coupling strength gives a higher steady state ergotropy. Intrigued by this feature, we ask what happens to the steady-state ergotropy if we increase the strength of the coherent drive. 
We find a complementarity between strengths of the nonlinear interaction, $g_2$, and the coherent drive, $F$, for obtaining the same steady-state ergotropy values.
Specifically we find that higher values of ergotropy are obtained in the scenarios when the value of $g_2$ is low and that of $F$ is high, or vice versa. The optimal value of steady-state ergotropy is found when both the nonlinear coupling strength and coherent drive strength are intermediate. Refer to Fig.~\ref{fig:new_max_g2_F_steady_heatmap} for this. Therefore,  higher values of the nonlinear coupling strength result in higher steady-state values, even for low values of coherent drive strength.

We found scenarios when nonlinearity in the coupling between battery and auxiliary assists the availability of maximum extractable energy in the steady state. Further, we found instances where the availability of maximum ergotropy and steady state ergotropy is facilitated at a lesser time, due to the presence of nonlinear interactions between auxiliary and battery.

The analysis presented thus far pertains to the scenario in which both single-photon and two-photon interactions jointly couple the battery to the cavity. However, the total interaction Hamiltonian comprising these couplings does not commute with the sum of the local Hamiltonians of the battery and the cavity. As a consequence, the energy change observed in the battery cannot be attributed solely to energy transferred from the cavity. Therefore, under this configuration, the cavity cannot be interpreted as a ``charger". For completeness, we also examine an alternative scenario in which the battery and the cavity interact via a combination of linear and nonlinear couplings that commute with the total local Hamiltonian, ensuring that the energy flow can be clearly attributed to the cavity acting as a charger.

\subsection{Interpreting the cavity as a charger}

Let us consider the nonlinear interaction between the cavity  and battery to be of the form
\begin{eqnarray}
    H_{AB}^{(2)} &=K\sqrt{\tilde{\omega}} g_2({a^{\dagger}}^ra^s \sigma^{+} + {a^{\dagger}}^sa^r \sigma^-),
\end{eqnarray}
where $g_2$ is the dimensionless nonlinear interaction strength, while $r$ and $s$ are integers such that $r+s>1$. In order to ensure that the commutator, $[H_{AB}^{(2)},H_A+H_B]=0$, the condition on the integers, $r$ and $s$ is given by $s-r=1$. We choose $s=2$ and $r=1$. Therefore, the total Hamiltonian that describes the coupling between the system and the cavity is given by
\begin{eqnarray}
    \widetilde{H}_{AB} &=& H_{AB}^{(1)}+ \widetilde{H}_{AB}^{(2)} \nonumber \\
    &=& \xi \left(g_1(a \sigma^{+} + a^{\dagger} \sigma^-) +   g_2({a^{\dagger}}a^2 \sigma^{+} + {a^{\dagger}}^2a \sigma^-) \right), \;\;\;\;\;
    \label{nonlin_com}
\end{eqnarray}
where $\xi =K\sqrt{\tilde{\omega}}$. 
The commutation, $[H_{AB},H_A+H_B]=0$, ensures that the interaction, $H_{AB}$, does not increase the sum of the local energies of the battery and the auxiliary. This implies that the change in energy of the battery is solely due to the flow of energy from the auxiliary. So the auxiliary behaves as a charger in this scenario. 

\begin{figure*}
\includegraphics[width=8.0cm]{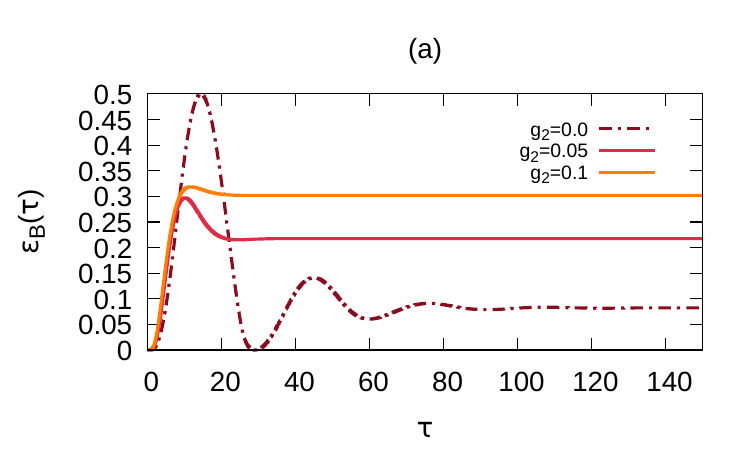}
\includegraphics[width=8.0cm]{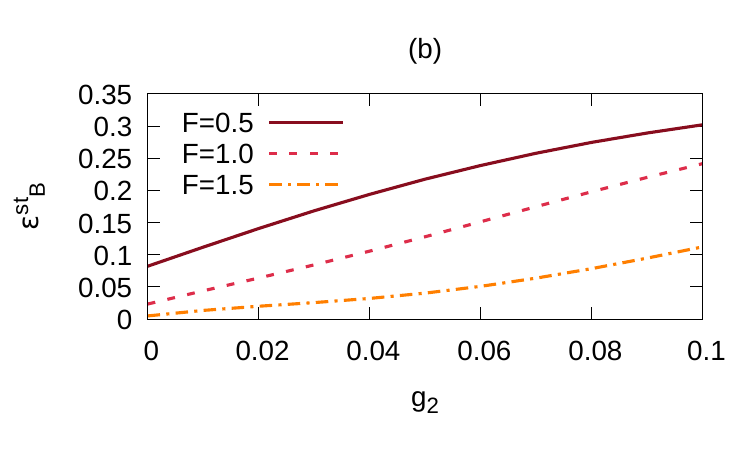}
\caption{\textbf{Panel (a):} Depiction of the maximum extractable energy versus time for different values of nonlinear interaction strength, $g_2$. The field strength corresponding to the coherent drive is $F = 0.5$, and the linear coupling strength is $g_1=0.1$.
Both in the main figure and inset, the quantities plotted along the horizontal axis is dimensionless, while ergotropy is units of $K$. 
\textbf{Panel (b):} The ergotropy of the battery, demonstrated as a function of the nonlinear interaction strength, $g_2$ for a different values of the driving field strength. The linear coupling strength is taken to be $g_1=0.1$. 
The quantities plotted along the horizontal axis is dimensionless, while ergotropy is in units of $K$. }
\label{nonlinear_charg_comm}
\end{figure*}

 The maximum extractable energy from quantum batteries when the battery is coupled to an auxiliary system via four-body interactions, as governed by the Hamiltonian presented in Eq.~\eqref{nonlin_com}.
The Hamiltonians $\Delta H_A$ and $H_{AB}$ (see Eqs.~\eqref{nonLinHamiltonian_HO_qubit} and~\eqref{nonlin_com}) contribute to the unitary evolution of the system, whereas the interaction of system A with the environment gives rise to the GKSL type dissipative terms. The resultant master equation, which follows from Eq.~\eqref{dissipator} is given by
\begin{eqnarray}
\label{comm_lindblad}
     \dot{\Tilde{\rho}}_{AB}(t) = &-&i [\sqrt{\tilde{\omega}} g_1(a \sigma^{+} + a  \sigma^-) \nonumber \\
        &+&\tilde{\omega} g_2({a^{\dagger}}a^2 \sigma^{+} + {a^{\dagger}}^2a \sigma^-) \nonumber \\
        &+& \sqrt{\tilde{\omega}} F(a^{\dagger}+a),\; \rho_{AB}(t)]  \nonumber \\
        &+&  J(\omega_0) (1+f) \Big[a \rho_{AB}(t) a^{\dagger}- \frac{1}{2} \{ a^{\dagger} a,\rho_{AB}(t) \} \Big] \nonumber \\
        &+& J(\omega_0) f \Big[a^{\dagger} \rho_{AB}(t) a - \frac{1}{2} \{ a a^{\dagger},\rho_{AB}(t) \} \Big], 
\end{eqnarray}
where $J(\omega_0)= \alpha \omega_0 \exp(-\omega_0/\omega_{max})$ and $f=f(\omega_{0},\beta)=[\exp(\beta \hbar \omega_{0})-1]^{-1}$ are the Ohmic spectral density
function and the Bose-Einstein distribution function respectively.
Here $\alpha$ represents the dimensionless interaction strength between the charger and the bosonic bath. For numerical purposes, $\alpha$ is chosen to be equal to $1$.

\subsubsection{Energy extraction when auxiliary is accessible}
\label{aux_access_comm}

This is the scenario in which the auxiliary is directly available for energy extraction. 
The master equation for the joint density matrix \( \rho_{AB}(t) \) is solved, following which the battery is traced out and the ergotropy of the auxiliary is calculated using Eq.~\eqref{erg}.
It is intuitive to find that the ergotropy of the auxiliary is higher than that of the battery in each of the linear and nonlinear coupling scenarios, since the auxiliary is directly driven by the coherent source of energy. 
The difference between the steady-state ergotropy of the charger and battery in the linear and nonlinear coupling scenarios are respectively given by $\Delta \varepsilon_{\text{lin}}=1.416$ and $\Delta \varepsilon_{\text{nonlin}}=1.008$.
Moreover, the steady-state ergotropy of the auxiliary in the nonlinearly coupled regime, given by $\varepsilon_A^{\text{st(nonlin)}}=1.310$, is considerably higher than its steady-state ergotropy in the linearly coupled situation, given by $\varepsilon_A^{\text{st(lin)}}=1.498$. This shows that nonlinear interaction between the battery and auxiliary also suffice to be useful if the auxiliary is accessible for energy extraction. These results correspond to the values of driving strength, $F=0.5$, linear coupling strength $g_1=0.1$, and nonlinear strength $g_2=0.1$.

\subsubsection{Energy extraction when auxiliary is not accessible}\label{aux_not_access_comm}

Here we analyze the situation when the cavity, i.e. the auxiliary, is not accessible and only the battery is available for energy extraction. After obtaining the solution of the master equation for the joint density matrix \( \rho_{AB}(t) \),  the charger is discarded and the ergotropy of the battery is computed using Eq.~\eqref{erg}. In the main plot of Fig.~\ref{nonlinear_charg_comm}(a), the ergotropy \( \varepsilon_B(\tau) \) of the battery is plotted as a function of time. It is observed that the oscillatory nature of \( \varepsilon_B(\tau) \) is suppressed in the presence of four-body interactions.
Moreover, our results demonstrate that nonlinear coupling between the charger and the battery offers a distinct advantage in the steady-state regime. Specifically, the steady-state ergotropy is higher for nonzero nonlinear coupling strengths compared to the purely linear case. Additionally, the system reaches its steady state more rapidly when nonlinear interactions are present. However, in the transient regime, no such advantage is observed, i.e. the maximum ergotropy attained during the transient period actually decreases with increasing nonlinearity.

These findings suggest that nonlinear interactions between the charger and the battery are particularly beneficial when continuous energy extraction at large timescales is desired. For our analysis, the linear coupling strength is fixed at \( g_1 = 0.1 \). We find an enhancement in steady-state ergotropy across both weak and strong nonlinear coupling regimes, i.e., when \( g_2 \) is close to $0$ and $0.1$, respectively.

We next examine the effect of varying the coherent drive strength on the steady-state ergotropy of the battery. In Fig.~\ref{nonlinear_charg_comm}(b), we present the steady-state ergotropy \( \varepsilon^{\text{st}}_B \) as a function of the nonlinear interaction strength \( g_2 \), for different values of the coherent drive amplitude \( F \). Our results reveal that, for moderate values of the drive strength within the range \( 0.5 \leq F \leq 1.5 \), the steady-state ergotropy increases as the nonlinear coupling strength \( g_2 \) increases. 
Furthermore, we observe that both the absolute value of the steady-state ergotropy and its rate of change with respect to \( g_2 \) diminish as the coherent drive strength \( F \) increases. This behavior highlights an interesting feature, viz. the maximum extractable energy in the steady state is enhanced when the coherent drive strength is relatively weak.

\section{Energy extraction assisted by anharmonic cavity in presence of environment}
\label{sec:anharmonic}

\begin{figure*} 
    \centering 
    \includegraphics[width=7.5cm]{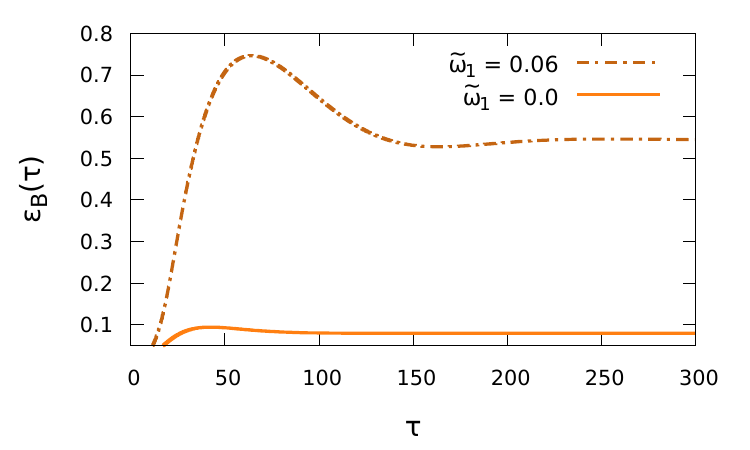} 
     \includegraphics[width=7.5cm]{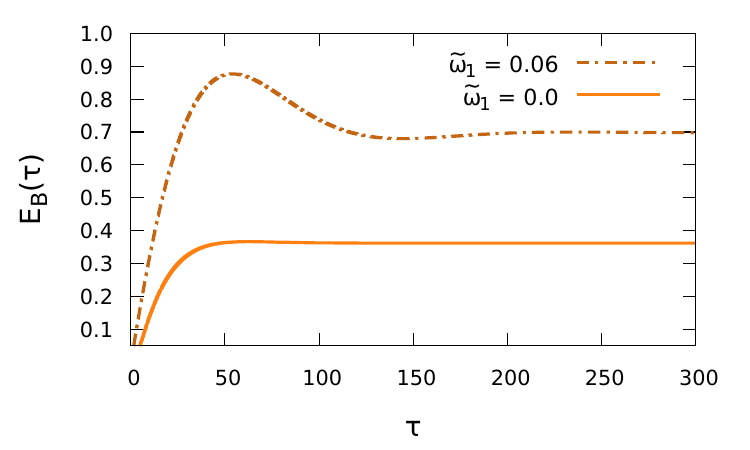}
    \caption{\textbf{Left panel:} Different curves correspond to the ergotropy versus interaction time for a fixed coherent drive $F=0.1$. Dotted-dashed and solid curves respectively represent the energies of the anharmonic and harmonic oscillators, as a function of the interaction time. The parameters chosen are $g = 0.1$, the anharmonic oscillator frequencies as $\tilde{\omega}_0 = 1.06$ and $\tilde{\omega}_1 = 0.06$. For the harmonic oscillator also,  $\tilde{\omega}_0 = 1.06$ is chosen. The quantities plotted along the horizontal axis is dimensionless, while ergotropy is in units of $K$. \textbf{Right panel:} Depiction of energy versus interaction time for the same set of values of the parameters. The quantities plotted along the horizontal axis is dimensionless, while energy is in units of $K$.}
    \label{ergotropy_AnHO_HO_diff_F0.1} 
\end{figure*}

In this section, we analyze the discharging of a quantum battery interacting with a bosonic Markovian environment through an auxiliary, which is considered to be a transmon.
We begin by considering the Hamiltonian of a quantum harmonic oscillator, which is obtained by quantizing the Hamiltonian of an $LC$ circuit, where  $L$ refers to inductance and $C$ is the capacitance. This Hamiltonian is usually applicable in superconductivity-based systems such as the Cooper-pair box ~\cite{nakamura1999coherent}. The quantized Hamiltonian is given by,
\begin{equation}
    \hat{H}_{HO} =\frac{\hat{Q}^2}{2C} + \frac{\hat{\Phi}^2}{2L},
\end{equation}
where $\hat{Q}$ refers to the charge variable and $\hat{\Phi}$ is the flux variable.
These operators satisfy the following commutation relation,
\begin{equation}
    [\hat{\Phi},\hat{Q}] =i\hbar,
\end{equation}
where $\hbar=h/2\pi$, and $h$ is the Planck's constant.
The Hamiltonian is written in a more recognizable form by introducing the reduced charge, $\hat{n}=\hat{Q}/2e$, where  2e is the charge content of a Cooper pair, and the phase $\hat{\phi}= 2\pi \hat{\Phi}/\phi_0$, where $\phi_0=h/2e$ is the flux quanta. Using this notation the Hamiltonian reduces to the form
\begin{equation}
    \hat{H}_{HO} = 4 E_c \hat{n}^2 + \frac{1}{2} E_L \hat{\phi}^2,
\end{equation}
where $E_c=e^2/2C$ is the charging energy and  $E_L= (\phi_0/2\pi)^2/L$ is the inductive energy.

Coupling two superconducting islands, or alternatively, an island and a ground through a Josephson junction ~\cite{PhysRevA.76.042319} is the standard set-up for a transmon. The effective Hamiltonian of the Cooper-pair box in the transmon regime can be written as ~\cite{devoret1997quantum,PhysRevA.76.042319, vool2017introduction} 
\begin{equation}
    \hat{H} = 4E_c\hat{n}^2 - E_J \cos \hat{\phi},
\end{equation}
where $E_J = I_0\phi_0/2\pi$ is the Josephson energy. 
The current $I$ and flux $\Phi$, across the Josephson junction are connected by the relation, $I=I_0 \sin(2\pi\Phi/\Phi_0)$.
Here we see that the phase is not in the form for that of a linear inductor. This is due to the presence of a Josephson junction.
We now express the charge and phase operators in terms of annihilation and creation operators of the transmon. The charge operator, $\hat{n}$ and phase operator, $\hat{\phi}$ are respectively given by
\begin{equation}
    \hat{n} =i \left(\frac{E_J}{32E_c}\right)^{1/4} (\hat{b}^{\dagger}-\hat{b}), \;\;
    \hat{\phi} =\left(\frac{2E_c}{E_J}\right)^{1/4} (\hat{b}^{\dagger}+\hat{b}). \;
\end{equation}
Here $\hat{b}=\sum_n \sqrt{n+1}|n\rangle \langle n+1|$ is the annihilation operator of the transmon, and unlike the harmonic oscillator, the energy modes are not evenly spaced. In the transmon regime, the Josephson and inductive energies hold the following inequality: $\beta=E_J/E_c>>1$. This allows one to take a Taylor expansion of $\cos\hat{\phi}$ and approximate the Hamiltonian to obtain
\begin{eqnarray}
        \hat{H} = &-&4E_c \sqrt{\frac{\beta}{32}} (\hat{b}^{\dagger}-\hat{b})^2 \nonumber\\
        &-& E_J \left(1-\frac{1}{2} \sqrt{\frac{2}{\beta}} (\hat{b}^{\dagger}+\hat{b})^2 +\frac{1}{12\beta} (\hat{b}^{\dagger}+\hat{b})^4 +...\right), \nonumber \\
        &\approx& \sqrt{8E_cE_J}\left(\hat{b}^{\dagger}\hat{b}+\frac{1}{2}\right)-E_J - \frac{E_c}{12}(\hat{b^{\dagger}}+\hat{b})^4.
\end{eqnarray}
Performing the rotating wave approximation and then defining $\omega '=\sqrt{8E_cE_J}$ and the anharmonicity of the transmon as $\delta=-E_c$, we obtain
\begin{equation}
    \hat{H} = \left(\omega ' +\frac{\delta}{2}\right) \hat{b}^{\dagger}\hat{b} + \frac{\delta}{2}(\hat{b}^{\dagger}\hat{b})^2.
\end{equation}

We consider the auxiliary to be an anharmonic oscillator represented by the transmon. The battery is a two-level system as considered in the previous section. The auxiliary interacts with a bosonic Markovian environment, $E$, through an interaction hamiltonian $H_{EA}$. The Hamiltonian of the transmon auxiliary is the same as $\hat{H}$. 
The total Hamiltonian involving the auxiliary and environment is given by
\begin{equation}
    \label{transmon_bath_hamiltonian}
        \tilde{H} =\tilde{H}_A + \tilde{H}_E +\tilde{H}_{EA},
\end{equation}
where $\tilde{H}_A= K \tilde{\omega} \tilde{\omega}_0 b^{\dagger}b - K \tilde{\omega}^2 \tilde{\omega}_1 (b^{\dagger}b)^2$, and $\tilde{H}_E=H_E$ as defined in Sec.~\ref{def}.
Here $b$($b^{\dagger}$) are the annihilation (creation) operators of the transmon auxiliary, having the unit of $1/\sqrt{\tilde{\omega}}$. The operators $a^{\dagger}_{\omega}(a_{\omega})$, having the unit of $1/\sqrt{\tilde{\omega}}$, represents the bosonic creation (annihilation) operators corresponding to the mode $\omega$ of the bath.
The notations, $K$, $\tilde{\omega}$ have their same meanings as in Sec.~\ref{prelim}, and the dimensionless quantities, $\tilde{\omega}_0$ and $\tilde{\omega}_1$ have their magnitudes equal to  $(\omega' +\delta/2)$ and $-\delta/2$ respectively.


\begin{figure*} 
    \centering 
    \includegraphics[width=7.5cm]{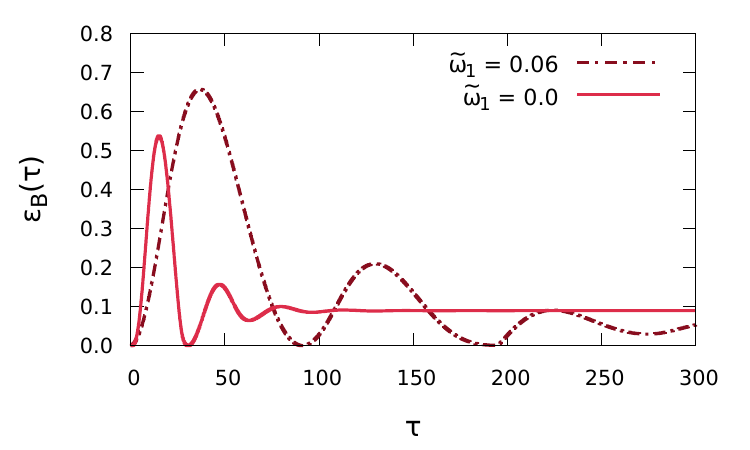} 
    \includegraphics[width=7.5cm]{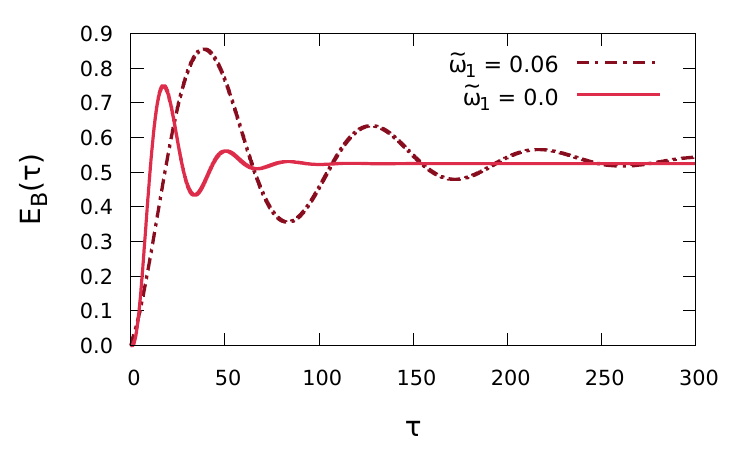}
    \caption{\textbf{Left panel:} Different curves correspond to the ergotropy versus interaction time for a fixed coherent drive $F=0.5$. Dotted-dashed and solid curves respectively represent the energies of the anharmonic and harmonic oscillators, as a function of the interaction time. The parameters chosen are $g = 0.1$, the anharmonic oscillator frequencies as $\tilde{\omega}_0 = 1.06$ and $\tilde{\omega}_1 = 0.06$. For the harmonic oscillator also,  $\tilde{\omega}_0 = 1.06$ is chosen. The quantities plotted along the horizontal axis is dimensionless, while ergotropy is in units of $K$. \textbf{Right panel:} Depiction of energy versus interaction time for the same set of values of the parameters.  Quantities plotted along both axes are dimensionless. The quantities plotted along the horizontal axis is dimensionless, while energy is in units of $K$.
} 
    \label{ergotropy_Anho_HO_specificF} 
\end{figure*}

The interaction between the system $A$ and the bath is given by
\begin{equation}
    \tilde{H}_{EA} = \int_0^{\omega_{max}} \sqrt{\tilde{\omega}} \hbar d\omega \tilde{h}(\omega) \left(b^{\dagger}a_{\omega}+b a^{\dagger}_{\omega}\right),
\end{equation}
where $\tilde{h}(\omega)$ denotes the strength of the coupling between the auxiliary $A$ and the environment.
The interaction between the auxiliary and the environment, i.e. $H_{EA}$, can be rewritten as a sum of the tensor product of Hermitian operators corresponding to the auxiliary and the bath, given by $H_{EA} = \sum_{i=1}^{2}A_i \otimes B_i$,
where $A_i$ and $B_i$ for $i=1$ to $2$ are the operators corresponding to the auxiliary $A$, and the bath $B$ respectively. In the case that we consider, the system operators are of the form
\begin{equation}
    \begin{split}
        A_1& = b^{\dagger}+b, \;\;\;
        A_2 =i (b^{\dagger}-b),
         \end{split} 
\end{equation}
The bath operators, denoted by $B_i$, for $i=1$ and $2$ are given by
\begin{equation}
    \begin{split}
        B_1 &= \int^{\omega_{max}}_0 d\omega h(\omega) \frac{(a_{\omega}+a^{\dagger}_{\omega})}{2} = \int^{\omega_{max}}_{-\omega_{max}} d\omega B_1(\omega), \\
        \text{where }&  B_1(\omega) = h(\omega) \frac{a_{\omega}}{2}\text{ and }  B_1(-\omega) = h(\omega) \frac{a_{\omega}^{\dagger}}{2},\\
        B_2 &= \int^{\omega_{max}}_0 d\omega h(\omega) \frac{i(a^{\dagger}_{\omega}-a_{\omega})}{2} = \int^{\omega_{max}}_{-\omega_{max}} d\omega B_2(\omega), \\
        \text{where }&  B_2(\omega) = -i h(\omega) \frac{a_{\omega}}{2}\text{ and }  B_2(-\omega) = i h(\omega) \frac{a_{\omega}^{\dagger}}{2}. 
    \end{split} 
\end{equation}

We now consider the system operators, $A_k$, and the eigenspectrum of $\tilde{H}_A$ being discrete with eigenvectors, $\ket{\psi_{\epsilon}}$, having eigenvalues, $\epsilon$, we define an operator $A_k(\omega)$, given by
\begin{equation}
\label{ak}
    A_k(\omega) = \sum_{\epsilon'-\epsilon =\omega} |\psi_{\epsilon} \rangle \langle \psi_{\epsilon}| A_k |\psi_{\epsilon'} \rangle\langle \psi_{\epsilon'}|,
\end{equation}
where $k=1$ or $2$.
The eigenvalues and eigenstates of the transmon Hamiltonian, $\tilde{H}_A$, are respectively given by
\begin{equation}
    \label{transmon_eigen_val_vect}
    \begin{split}
    \epsilon_n = \tilde{\omega}_0 n -\tilde{\omega}_1 n^2, \\
    |\psi_{\epsilon} \rangle = \frac{b^{\dagger n}}{\sqrt{n!}} | \psi_0 \rangle,
    \end{split}
\end{equation}
where $n$ denotes the $n^{\text{th}}$ level of the transmon.
So using equations~\eqref{ak} and~\eqref{transmon_eigen_val_vect}, for $n=n'+1$,  using the commutation relation of the annihilation and creation operators, we obtain
\begin{equation}
\begin{split}
    A_1(\omega_{n'}) = \sqrt{n'+1} |\psi_{\epsilon_{n'+1}} \rangle \langle \psi_{\epsilon_{n'}}|= T_{\uparrow n'}.
\end{split}
\end{equation}
where the dimensionless quantity, $ \omega_{n'}= \epsilon_{n'}-\epsilon_{n'+1} =-\omega_0+2\omega_1n'+\omega_1$.
Whereas for $n+1=n'$, similarly we arrive at the expression
\begin{equation}
\begin{split}
    A_1(\omega_{n}) = \sqrt{n+1}|\psi_{\epsilon_{n}} \rangle \langle \psi_{\epsilon_{n+1}}|= T_{\downarrow n}
\end{split}
\end{equation}
where the dimensionless quantity, $\omega_{n}= \epsilon_{n+1}-\epsilon_{n}=\tilde{\omega}_0-2\omega_1n-\tilde{\omega}_1$. 
Therefore the energy differences associated with the allowed transitions, $n \rightarrow n+1$ and $n+1 \rightarrow n$ are $\omega_{n}=-\tilde{\omega}_0+2\omega_1n+\tilde{\omega}_1 = \omega_{\uparrow n}=-\omega_{\downarrow n}$ and $\omega_{n}=\tilde{\omega}_0-2\tilde{\omega}_1n-\tilde{\omega}_1 = \omega_{\downarrow n}$ respectively.
Similarly, the operators corresponding to $A_2$, which are associated with the transitions $n \rightarrow n+1$ and $n+1 \rightarrow n$   are given by
\begin{equation}
\begin{split}
     A_2(\omega_{\uparrow n}) &= -i \sqrt{n+1}|\psi_{\epsilon_{n}} \rangle \langle \psi_{\epsilon_{n+1}}|= T_{\downarrow n} \;\; \text{and}\\
     A_2(\omega_{\downarrow n}) &= i \sqrt{n+1} |\psi_{\epsilon_{n+1}} \rangle \langle \psi_{\epsilon_{n}}|= T_{\uparrow n}
\end{split}
\end{equation}
respectively.

Now, to complete the description of the evolution of the auxiliary and the battery, we seek the decay rates of the GKSL master equation~\eqref{lindblad2}. Following the exposition in ~\cite{rivasarxivcitation}, we see that the decay rates for each transition corresponding to the energy difference $\omega_n$ are given by,
\begin{equation}
    \gamma_{ij}(\omega_n) = 2\pi \Tr[B_i(\omega_n)B_j \rho_\mathcal{B}], \;\;\; \forall i,j,
\end{equation}
where $\rho_\mathcal{B}$ is the density matrix of the bath, which is considered to be a thermal state comprising of infinite number of harmonic oscillators.

\begin{figure*} 
    \centering 
    \includegraphics[width=7.5cm]{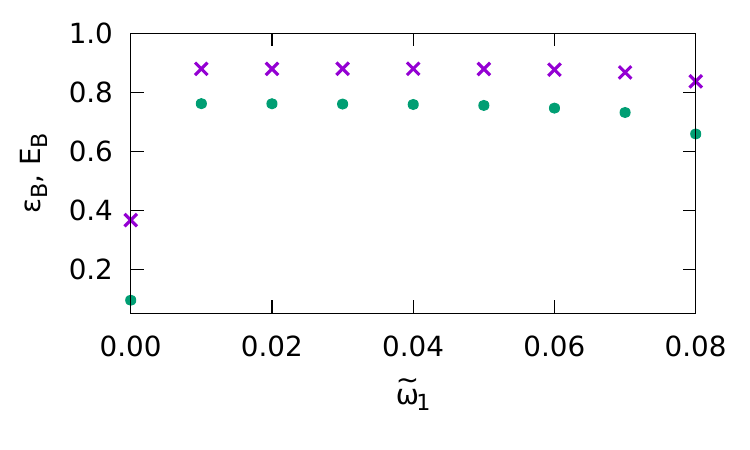} 
    \includegraphics[width=7.5cm]{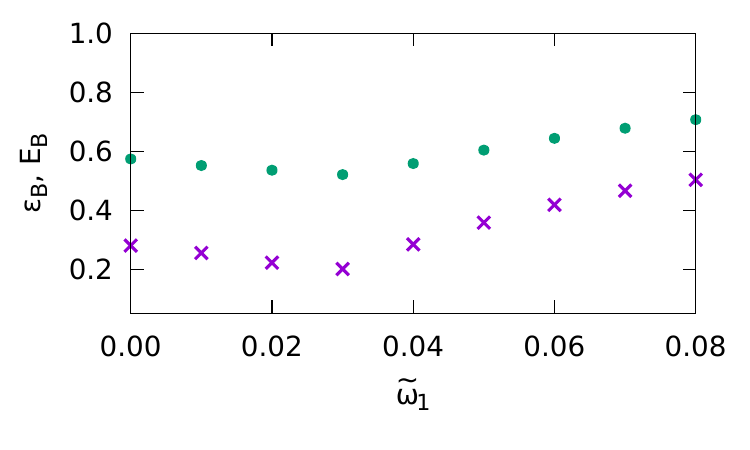}
    \caption{\textbf{Left panel:} Demonstration of the maximum ergotropy and maximum energy obtainable as a function of the anharmonicity parameter, $\tilde{\omega}_1$.
    Teal circles denote the value of maximum ergotropy, $\varepsilon_{B}(\tau)$ attained for different values of anharmonicity. The Purple crosses denote the same quantity, but for the case of the energy $E_{B}(\tau)$. Here we have obtained the results by setting $g_1 = 0.1$, $\gamma = 1.0$, $F=0.1$, $\tilde{\alpha}=1.0$ and we have set inverse temperature $\beta = 1$. The quantities plotted along the horizontal axis is dimensionless, while energy and ergotropy are in units of $K$. \textbf{Right panel:} The same quantities are plotted for the parameter value \( \tilde{\alpha} = 0.5 \), while all other parameters are kept identical to those used in the left panel.} 
    \label{newF_vs_tau_anharmonic} 
\end{figure*}

Substituting the bath operators from the interaction term, $\tilde{H}_{EA}$, the decay rate for $i,j=1$ is given by
\begin{equation}
\begin{split}
\label{gamma}
    \gamma_{11}(\omega_n) &= 2\pi \Tr[B_1(\omega_n)B_1 \rho_\mathcal{B}]\\
    &= 2 \pi  \int^{\omega_{max}}_0 \tilde{h}(\omega)\tilde{h}(\omega_n) \Tr\left[\frac{a_{\omega_n}(a_{\omega}+a^{\dagger}_{\omega})}{4}\rho_\mathcal{B}\right]  d\omega,
\end{split} 
\end{equation}
where $\omega_{max}$ denotes the cutoff frequency of the bath. This cutoff frequency is set to be sufficiently high so that the memory time of the bath $\sim \omega_{max}^{-1}$, is negligibly small.
Explicitly calculating~\eqref{gamma}, we obtain 
\begin{equation}
    \gamma_{11}(\omega_n) = \frac{\pi}{2} \tilde{h}^2(\omega_n)[\Bar{n}(\omega_n)+1],
\end{equation}
where $J(\omega_n)=\tilde{h}^2(\omega_n)$, is the spectral density of the bath at frequency $\omega_n$, $\Bar{n}(\omega_n)$ is the mean number of bosons in the thermal state $\rho_\mathcal{B}$ of the bath with frequency $\omega_n$, which is given by $\Bar{n}(\omega_n) = \left[e^{\beta \omega_n/k_B}-1\right]^{-1}$, where
$\beta$ denotes the inverse temperature of the bath. Upon doing a similar calculation of the other elements of the decay rates, we obtain
\begin{equation}
\begin{split}
       \mathbf{\gamma}(\omega_{\downarrow n}) &=  \frac{\pi}{2} \tilde{h}^2(\omega_n)[\Bar{n}(\omega_{\downarrow n})+1] \begin{pmatrix}1 & -i \\
i & 1 \\
\end{pmatrix} \;\;\text{and}  \\
       \mathbf{\gamma}(-\omega_{\downarrow n}) &=  \frac{\pi}{2} \tilde{h}^2(\omega_n)\Bar{n}(\omega_{\downarrow n}) \begin{pmatrix}1 & i \\
-i & 1 \\
\end{pmatrix}.
\end{split}
\end{equation} 
Now that we have everything we need for the master equation that describes the dynamics of the battery and auxiliary, the GKSL master equation corresponding to Eq~\eqref{lindblad} for the battery and auxiliary is given by
\begin{equation}
\label{AnHO_QB_ME}
\begin{split}
    \dot{\Tilde{\rho}}_{AB}(t) = &-i [\sqrt{\tilde{\omega}}g(b \sigma^{+} + b^{\dagger} \sigma^-) \,+\sqrt{\tilde{\omega}} F(b^{\dagger}+b),\; \rho_{AB}(t)] \\
    &+ \sum_n  \xi_n \left[T_{\downarrow n}\rho_{AB}(t)T_{\uparrow n}-\frac{1}{2}\{T_{\downarrow n}T_{\uparrow n},\rho_{AB}(t)\} \right]\\
    &+ \sum_n  \xi_n \left[T_{\uparrow n}\rho_{AB}(t)T_{\downarrow n}-\frac{1}{2}\{T_{\uparrow n}T_{\downarrow n},\rho_{AB}(t)\} \right], 
\end{split}
\end{equation}
where $\xi_n=2\pi J(\omega_n)[\Bar{n}(\omega_n)+1]$. For harmonic oscillator baths, $\tilde{\omega} \tilde{h}^{2}(\omega_n)=J(\omega_n)$, where $J(\omega_n)$ represents the spectral density function of the bath. The quantity $J(\omega_n)= \tilde{\alpha} \omega_n \exp(-\omega_n/\omega_{max})$. Here $\tilde{\alpha}$ represents the dimensionless interaction strength between the anharmonic-oscillator and the bosonic bath. 
Here we have taken the interaction between the auxiliary and battery to be linear. 
In contrast to the harmonic oscillator case, in this scenario, we see that the annihilation and creation operators in the dissipative part of the GKSL master equation are replaced by $n^{\text{th}}$ site transition operators, and we sum over all possible transitions of the anharmonic oscillator.
The GKSL master equation for a multilevel transmon immersed in a bosonic Markovian bath, which is derived in Eq.~\eqref{AnHO_QB_ME}, is novel and is hitherto unexplored in the literature.\\

\subsection{Does anharmonicity help in discharging?}
Here we present the details of implementation of the dynamics of a two-level system as battery and an anharmonic auxiliary, and we will study the resulting effect on maximum extractable energy from the battery using unitary operations. 
The dynamics of the anharmonic auxiliary and the two-level battery (system $B$) is governed by the master equation described in Eq.~\eqref{AnHO_QB_ME}. 
We solve the master equation numerically using the Runge-Kutta method, to obtain $\rho_{AB}$ as a function of interaction time, following which we discard the system $A$, and consider the system $B$.

In the case where there is no coherent driving, i.e. $F=0$,  our findings are in agreement with the expectation in that there is no ergotropy left on $B$ while having a finite $E_B(\tau)$. Switching on the coherent drive $F$ and increasing its value results in nonzero values of $\varepsilon_B(\tau)$ that exhibit an oscillatory behavior that is a result of a competing effect between the dissipative and the coherent drive.


Fig.~\ref{ergotropy_AnHO_HO_diff_F0.1}-(a) depicts the maximum extractable energy using unitary operations, i.e. the ergotropy, that is available from the quantum battery, $B$, as a function of the interaction time, $\tau$, if the auxiliary is considered to be a anharmonic oscillator. We find that the amount of maximum extractable energy at all times is greater in the case where the auxiliary is an anharmonic oscillator, than that of the harmonic oscillator case. Fig.~\ref{ergotropy_AnHO_HO_diff_F0.1}-(a) is plotted for the values of the anharmonicity parameter, $\tilde{\omega}_1=0.06$, and the coherent driving strength, $F=0.1$. This feature is also prevalent if we consider the energy of the battery, $B$, as a function of time. This implies that the amount of energy stored in the process of charging at any time of interaction, is higher in the case of anharmonic oscillator auxiliary than that for the harmonic oscillator case. Refer to Fig.~\ref{ergotropy_AnHO_HO_diff_F0.1}-(b) for this. 
So using transmon as an auxiliary is advantageous over using a simple harmonic oscillator in the process of charging or extracting maximal amount of energy from a quantum battery at all timescales, for a certain coherent drive.

For higher values of coherent drive strength, we find an advantage in the maximum ergotropy achieved over the interaction time, by using an anharmonic auxiliary, over the harmonic one. However, the oscillatory behavior is increased for the anharmonic oscillator, and it takes a longer time to achieve a steady state.  Refer to Fig.~\ref{ergotropy_Anho_HO_specificF}-(a). If we increase the value of the driving amplitude further, there is a drop in ergotropy value when compared to the harmonic oscillator case. We identify a range of values of the coherent drive strength for which there is an advantage in using an anharmonic auxiliary instead of a harmonic one.
Therefore, if there is a requirement of high values of ergotropy available at comparatively low timescales, then anharmonic auxiliarys prove to be advantageous over harmonic ones. Qualitatively similar features are also present in the plot of energy extracted versus the interaction time. The maximum energy obtained over time is higher if we use an anharmonic auxiliary, however it takes a longer time to achieve a steady state. Refer to Fig.~\ref{ergotropy_Anho_HO_specificF}-(b) for this.

\begin{figure} 
    \centering 
    \includegraphics[width=7.8cm]{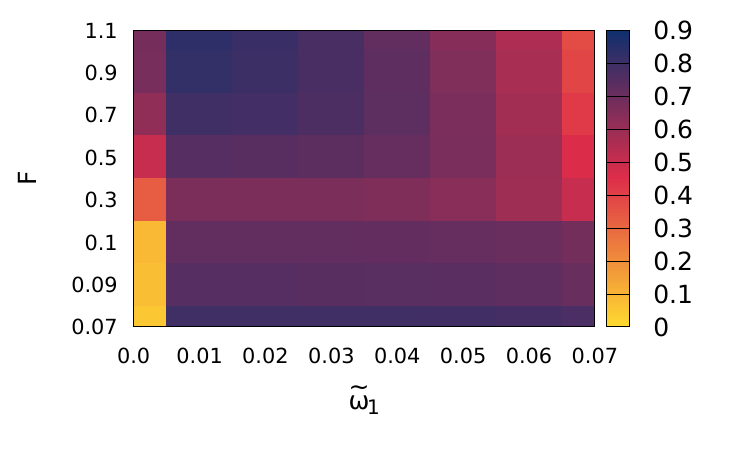} 
    \caption{Depiction of the maximum ergotropy, where the maximization is performed over time, obtained for different values of the coherent drive intensity $F$ along the vertical axis, and the anharmonic strength, $\tilde{\omega}_1$, along the horizontal axis. The quantities plotted along the horizontal and vertical axes are dimensionless, while ergotropy is in units of $K$.} 
    \label{fig:newAnHO_max_g1_F_heatmap} 
\end{figure} 

We independently maximize the ergotropy and stored energy over the timescale required for the system to reach a steady state and plot the resulting values as functions of the anharmonicity strength \( \tilde{\omega}_1 \). It is observed that, for specific values of the coherent drive, there exists a regime of \( \tilde{\omega}_1 \) where the maximum values of ergotropy and energy are higher when an anharmonic auxiliary is used, compared to a harmonic one. This behavior is illustrated in the left panel of Fig.~\ref{newF_vs_tau_anharmonic}. Furthermore, this advantage is influenced by both the parameter \( \tilde{\alpha} \), which governs the dissipation rate of the auxiliary system, and the temperature of the environment. In particular, there exist values of these parameters for which the use of an anharmonic auxiliary does not result in any improvement in the maximum ergotropy or stored energy over the harmonic case. For instance,  as shown in the right panel of Fig.~\ref{newF_vs_tau_anharmonic}, corresponding to \( \tilde{\alpha} = 0.5 \), the anharmonic auxiliary yields superior performance only when the anharmonicity strength exceeds \( 0.03 \). However, no universal trend is found in how the dissipation rate and bath temperature affect the ergotropy in the nonlinear regime.

Furthermore, this advantage is influenced by both the parameter \( \tilde{\alpha} \)—which governs the dissipation rate of the auxiliary system—and the temperature of the environment. In particular, there exist values of these parameters for which the use of an anharmonic auxiliary does not result in any improvement in the maximum ergotropy or stored energy over the harmonic case. For example, as shown in the right panel of Fig.~\ref{newF_vs_tau_anharmonic}, corresponding to \( \tilde{\alpha} = 0.5 \), the anharmonic auxiliary yields superior performance only when the anharmonicity strength exceeds \( 0.03 \). However, no universal trend is found in how the dissipation rate and bath temperature affect the ergotropy in the nonlinear regime, indicating a complex interplay between system parameters and nonlinearity.

Furthermore, this advantage is influenced by both the parameter \( \tilde{\alpha} \)—which governs the dissipation rate of the auxiliary system—and the temperature of the environment. In particular, there exist values of these parameters for which the use of an anharmonic auxiliary does not result in any improvement in the maximum ergotropy or stored energy over the harmonic case. For example, as shown in the right panel of Fig.~\ref{newF_vs_tau_anharmonic}, corresponding to \( \tilde{\alpha} = 0.5 \), the anharmonic auxiliary yields superior performance only when the anharmonicity strength exceeds \( 0.03 \). However, no universal trend is found in how the dissipation rate and bath temperature affect the ergotropy in the nonlinear regime, indicating a complex interplay between system parameters and nonlinearity.

Intrigued by the complementarity that we obtained in the case of nonlinearity assisted charging, we try to find if any such feature also exists here.
The heat map in Fig.~\ref{fig:newAnHO_max_g1_F_heatmap} demonstrates the maximum  ergotropy attained over time, for different values of the parameters, $F$ and $\tilde{\omega}_1$. We find that, for a lower values of the coherent drive, the entire range of anharmonic strength, $\tilde{\omega}_1$, from $0.01$ to $0.07$, proves to be beneficial in the availability of maximal extractable energy. In the case when the anharmonicity strength is $0$, the maximum ergotropy is almost negligible for small values of coherent drive, and it increases slowly on increasing the driving strength. Moreover, at lower values of $\tilde{\omega}_1$ $\sim 0.01$, high values of the coherent drive $\sim 1.0$ is required to attain the maximum ergotropy. Therefore we find that the anharmonicity in the auxiliary is potentially useful in extracting energy maximally from a battery even for a small coherent drive. \\

\section{Conclusion}
\label{concl}
We have analyzed the performance of a quantum battery, focusing on both energy storage and extraction, by incorporating nonlinear interactions between a qubit battery and a single-mode cavity within a Dicke model framework, employing an open-system approach. 
Specifically, we have studied two varieties of nonlinearities in the set-ups: non-linearly coupled cavity and battery, 
and cavity comprising an anharmonic oscillator which is linearly coupled to a two-level battery.
In both the scenarios, the battery is not in direct contact of the environment and only the cavity interacts with the environment and is 
acted on by a coherent drive. Before the external drive and the interaction with the environment is switched on, we began with the ground states of both the cavity and battery. The cavity is also referred to as the auxiliary in our analysis.

In the case of nonlinearity in the coupling between auxiliary and battery, we found that there is an enhancement in the steady-state ergotropy over the linear case. We also found instances where the ergotropy in the transient state is zero in the linear case, but 
is positive for all non-zero values of 
the 
nonlinear interaction strength. Further, we found that the times of attainment of maximum and steady-state ergotropies, where the maximization is over time, get reduced if we use nonlinear interaction between auxiliary and battery. 
We also found a complementarity between strengths of the nonlinear interaction and the coherent drive for obtaining the same steady-state ergotropy values.
For completeness, we also identify a class of nonlinear interactions under which the cavity can be legitimately regarded as a ``charger,'' and demonstrate that such nonlinear couplings offer an advantage over their linear counterparts in enhancing battery performance.

We obtained the exact form of the master equation of the auxiliary-battery system, when the auxiliary is  an anharmonic oscillator, and is connected to a bosonic Markovian environment. Considering the auxiliary to be an anharmonic oscillator, for example a transmon, we found that there is a significant advantage in the maximum ergotropy over the linear setup, though the maximum is attained at a slightly later time for the anharmonic case. 
We also identified instances where the anharmonicity-assisted charging proves to be beneficial over the harmonic case, over all timescales.
We separately maximized the ergotropy and energy, over the timescale in which the system attains a steady state, and 
found that for certain values of the coherent drive, there is a specific region of the anharmonicity strength where the maximum values of ergotropy and energy are higher in the case of an anharmonic auxiliary than 
when 
using a harmonic one. Moreover, we observed a specific region of the parameter space of anharmonic strength and coherent drive where the ergotropy values are maximum.

\acknowledgements
A.B. acknowledges support from ‘INFOSYS scholarship for senior students’ at Harish Chnadra Research Institute, India.
    
\bibliography{Bibliography}

\begin{thebibliography}{77}%
\makeatletter
\providecommand \@ifxundefined [1]{%
 \@ifx{#1\undefined}
}%
\providecommand \@ifnum [1]{%
 \ifnum #1\expandafter \@firstoftwo
 \else \expandafter \@secondoftwo
 \fi
}%
\providecommand \@ifx [1]{%
 \ifx #1\expandafter \@firstoftwo
 \else \expandafter \@secondoftwo
 \fi
}%
\providecommand \natexlab [1]{#1}%
\providecommand \enquote  [1]{``#1''}%
\providecommand \bibnamefont  [1]{#1}%
\providecommand \bibfnamefont [1]{#1}%
\providecommand \citenamefont [1]{#1}%
\providecommand \href@noop [0]{\@secondoftwo}%
\providecommand \href [0]{\begingroup \@sanitize@url \@href}%
\providecommand \@href[1]{\@@startlink{#1}\@@href}%
\providecommand \@@href[1]{\endgroup#1\@@endlink}%
\providecommand \@sanitize@url [0]{\catcode `\\12\catcode `\$12\catcode `\&12\catcode `\#12\catcode `\^12\catcode `\_12\catcode `\%12\relax}%
\providecommand \@@startlink[1]{}%
\providecommand \@@endlink[0]{}%
\providecommand \url  [0]{\begingroup\@sanitize@url \@url }%
\providecommand \@url [1]{\endgroup\@href {#1}{\urlprefix }}%
\providecommand \urlprefix  [0]{URL }%
\providecommand \Eprint [0]{\href }%
\providecommand \doibase [0]{http://dx.doi.org/}%
\providecommand \selectlanguage [0]{\@gobble}%
\providecommand \bibinfo  [0]{\@secondoftwo}%
\providecommand \bibfield  [0]{\@secondoftwo}%
\providecommand \translation [1]{[#1]}%
\providecommand \BibitemOpen [0]{}%
\providecommand \bibitemStop [0]{}%
\providecommand \bibitemNoStop [0]{.\EOS\space}%
\providecommand \EOS [0]{\spacefactor3000\relax}%
\providecommand \BibitemShut  [1]{\csname bibitem#1\endcsname}%
\let\auto@bib@innerbib\@empty
\bibitem [{\citenamefont {Bardeen}\ and\ \citenamefont {Brattain}(1949)}]{bardeen1949physical}%
  \BibitemOpen
  \bibfield  {author} {\bibinfo {author} {\bibfnamefont {J.}~\bibnamefont {Bardeen}}\ and\ \bibinfo {author} {\bibfnamefont {W.~H.}\ \bibnamefont {Brattain}},\ }\bibfield  {title} {\enquote {\bibinfo {title} {Physical principles involved in transistor action},}\ }\href {\doibase 10.1103/PhysRev.75.1208} {\bibfield  {journal} {\bibinfo  {journal} {Phys. Rev.}\ }\textbf {\bibinfo {volume} {75}},\ \bibinfo {pages} {1208--1225} (\bibinfo {year} {1949})}\BibitemShut {NoStop}%
\bibitem [{\citenamefont {Pearson}\ and\ \citenamefont {Brattain}(1955)}]{pearson1955history}%
  \BibitemOpen
  \bibfield  {author} {\bibinfo {author} {\bibfnamefont {G.~L.}\ \bibnamefont {Pearson}}\ and\ \bibinfo {author} {\bibfnamefont {W.~H.}\ \bibnamefont {Brattain}},\ }\bibfield  {title} {\enquote {\bibinfo {title} {History of semiconductor research},}\ }\href {https://api.semanticscholar.org/CorpusID:51634231} {\bibfield  {journal} {\bibinfo  {journal} {Proceedings of the IRE}\ }\textbf {\bibinfo {volume} {43}},\ \bibinfo {pages} {1794--1806} (\bibinfo {year} {1955})}\BibitemShut {NoStop}%
\bibitem [{\citenamefont {Bromberg}(1988)}]{bromberg1988birth}%
  \BibitemOpen
  \bibfield  {author} {\bibinfo {author} {\bibfnamefont {J.~L.}\ \bibnamefont {Bromberg}},\ }\bibfield  {title} {\enquote {\bibinfo {title} {{The Birth of the Laser}},}\ }\href {\doibase 10.1063/1.881155} {\bibfield  {journal} {\bibinfo  {journal} {Physics Today}\ }\textbf {\bibinfo {volume} {41}},\ \bibinfo {pages} {26--33} (\bibinfo {year} {1988})}\BibitemShut {NoStop}%
\bibitem [{\citenamefont {Preskill}(2018)}]{Preskill2018quantumcomputingin}%
  \BibitemOpen
  \bibfield  {author} {\bibinfo {author} {\bibfnamefont {J.}~\bibnamefont {Preskill}},\ }\bibfield  {title} {\enquote {\bibinfo {title} {Quantum {C}omputing in the {NISQ} era and beyond},}\ }\href {\doibase 10.22331/q-2018-08-06-79} {\bibfield  {journal} {\bibinfo  {journal} {{Quantum}}\ }\textbf {\bibinfo {volume} {2}},\ \bibinfo {pages} {79} (\bibinfo {year} {2018})}\BibitemShut {NoStop}%
\bibitem [{\citenamefont {Gisin}\ and\ \citenamefont {Thew}(2007)}]{gisin2007quantum}%
  \BibitemOpen
  \bibfield  {author} {\bibinfo {author} {\bibfnamefont {N.}~\bibnamefont {Gisin}}\ and\ \bibinfo {author} {\bibfnamefont {R.}~\bibnamefont {Thew}},\ }\bibfield  {title} {\enquote {\bibinfo {title} {Quantum communication},}\ }\href {\doibase 10.1038/nphoton.2007.22} {\bibfield  {journal} {\bibinfo  {journal} {Nature photonics}\ }\textbf {\bibinfo {volume} {1}},\ \bibinfo {pages} {165--171} (\bibinfo {year} {2007})}\BibitemShut {NoStop}%
\bibitem [{\citenamefont {Blatt}\ and\ \citenamefont {Roos}(2012)}]{blatt2012quantum}%
  \BibitemOpen
  \bibfield  {author} {\bibinfo {author} {\bibfnamefont {R.}~\bibnamefont {Blatt}}\ and\ \bibinfo {author} {\bibfnamefont {C.~F.}\ \bibnamefont {Roos}},\ }\bibfield  {title} {\enquote {\bibinfo {title} {Quantum simulations with trapped ions},}\ }\href {\doibase 10.1038/nphys2252} {\bibfield  {journal} {\bibinfo  {journal} {Nature Physics}\ }\textbf {\bibinfo {volume} {8}},\ \bibinfo {pages} {277--284} (\bibinfo {year} {2012})}\BibitemShut {NoStop}%
\bibitem [{\citenamefont {Alicki}(1979)}]{alicki1979quantum}%
  \BibitemOpen
  \bibfield  {author} {\bibinfo {author} {\bibfnamefont {R.}~\bibnamefont {Alicki}},\ }\bibfield  {title} {\enquote {\bibinfo {title} {The quantum open system as a model of the heat engine},}\ }\href {\doibase 10.1088/0305-4470/12/5/007} {\bibfield  {journal} {\bibinfo  {journal} {Journal of Physics A: Mathematical and General}\ }\textbf {\bibinfo {volume} {12}},\ \bibinfo {pages} {L103} (\bibinfo {year} {1979})}\BibitemShut {NoStop}%
\bibitem [{\citenamefont {Deffner}\ and\ \citenamefont {Campbell}(2019)}]{deffner2019quantum}%
  \BibitemOpen
  \bibfield  {author} {\bibinfo {author} {\bibfnamefont {S.}~\bibnamefont {Deffner}}\ and\ \bibinfo {author} {\bibfnamefont {S.}~\bibnamefont {Campbell}},\ }\bibfield  {title} {\enquote {\bibinfo {title} {Quantum thermodynamics},}\ }\href@noop {} {\bibfield  {journal} {\bibinfo  {journal} {Morgan \& Claypool Publishers}\ }\textbf {\bibinfo {volume} {10}},\ \bibinfo {pages} {2053--2571} (\bibinfo {year} {2019})}\BibitemShut {NoStop}%
\bibitem [{\citenamefont {Millen}\ and\ \citenamefont {Xuereb}(2016)}]{Millen_2016}%
  \BibitemOpen
  \bibfield  {author} {\bibinfo {author} {\bibfnamefont {J.}~\bibnamefont {Millen}}\ and\ \bibinfo {author} {\bibfnamefont {A.}~\bibnamefont {Xuereb}},\ }\bibfield  {title} {\enquote {\bibinfo {title} {Perspective on quantum thermodynamics},}\ }\href {\doibase 10.1088/1367-2630/18/1/011002} {\bibfield  {journal} {\bibinfo  {journal} {New J. Phys.}\ }\textbf {\bibinfo {volume} {18}},\ \bibinfo {pages} {011002} (\bibinfo {year} {2016})}\BibitemShut {NoStop}%
\bibitem [{\citenamefont {Scovil}\ and\ \citenamefont {Schulz-DuBois}(1959)}]{PhysRevLett.2.262}%
  \BibitemOpen
  \bibfield  {author} {\bibinfo {author} {\bibfnamefont {H.~E.~D.}\ \bibnamefont {Scovil}}\ and\ \bibinfo {author} {\bibfnamefont {E.~O.}\ \bibnamefont {Schulz-DuBois}},\ }\bibfield  {title} {\enquote {\bibinfo {title} {Three-level masers as heat engines},}\ }\href {\doibase 10.1103/PhysRevLett.2.262} {\bibfield  {journal} {\bibinfo  {journal} {Phys. Rev. Lett.}\ }\textbf {\bibinfo {volume} {2}},\ \bibinfo {pages} {262--263} (\bibinfo {year} {1959})}\BibitemShut {NoStop}%
\bibitem [{\citenamefont {Kosloff}(1984)}]{kosloff1984quantum}%
  \BibitemOpen
  \bibfield  {author} {\bibinfo {author} {\bibfnamefont {R.}~\bibnamefont {Kosloff}},\ }\bibfield  {title} {\enquote {\bibinfo {title} {A quantum mechanical open system as a model of a heat engine},}\ }\href {\doibase 10.1063/1.446862} {\bibfield  {journal} {\bibinfo  {journal} {The Journal of chemical physics}\ }\textbf {\bibinfo {volume} {80}},\ \bibinfo {pages} {1625--1631} (\bibinfo {year} {1984})}\BibitemShut {NoStop}%
\bibitem [{\citenamefont {Saha}\ \emph {et~al.}(2023)\citenamefont {Saha}, \citenamefont {Bhattacharyya}, \citenamefont {Sen},\ and\ \citenamefont {Sen}}]{saha2023}%
  \BibitemOpen
  \bibfield  {author} {\bibinfo {author} {\bibfnamefont {D.}~\bibnamefont {Saha}}, \bibinfo {author} {\bibfnamefont {A.}~\bibnamefont {Bhattacharyya}}, \bibinfo {author} {\bibfnamefont {K.}~\bibnamefont {Sen}}, \ and\ \bibinfo {author} {\bibfnamefont {U.}~\bibnamefont {Sen}},\ }\bibfield  {title} {\enquote {\bibinfo {title} {Harnessing energy extracted from heat engines to charge quantum batteries},}\ }\href {https://arxiv.org/abs/2309.15634} {\bibfield  {journal} {\bibinfo  {journal} {arXiv:2309.15634}\ } (\bibinfo {year} {2023})}\BibitemShut {NoStop}%
\bibitem [{\citenamefont {Linden}\ \emph {et~al.}(2010)\citenamefont {Linden}, \citenamefont {Popescu},\ and\ \citenamefont {Skrzypczyk}}]{popescu}%
  \BibitemOpen
  \bibfield  {author} {\bibinfo {author} {\bibfnamefont {Noah}\ \bibnamefont {Linden}}, \bibinfo {author} {\bibfnamefont {Sandu}\ \bibnamefont {Popescu}}, \ and\ \bibinfo {author} {\bibfnamefont {Paul}\ \bibnamefont {Skrzypczyk}},\ }\bibfield  {title} {\enquote {\bibinfo {title} {How small can thermal machines be? the smallest possible refrigerator},}\ }\href {\doibase 10.1103/PhysRevLett.105.130401} {\bibfield  {journal} {\bibinfo  {journal} {Phys. Rev. Lett.}\ }\textbf {\bibinfo {volume} {105}},\ \bibinfo {pages} {130401} (\bibinfo {year} {2010})}\BibitemShut {NoStop}%
\bibitem [{\citenamefont {Clivaz}\ \emph {et~al.}(2019)\citenamefont {Clivaz}, \citenamefont {Silva}, \citenamefont {Haack}, \citenamefont {Brask}, \citenamefont {Brunner},\ and\ \citenamefont {Huber}}]{PhysRevLett.123.170605}%
  \BibitemOpen
  \bibfield  {author} {\bibinfo {author} {\bibfnamefont {F.}~\bibnamefont {Clivaz}}, \bibinfo {author} {\bibfnamefont {R.}~\bibnamefont {Silva}}, \bibinfo {author} {\bibfnamefont {G.}~\bibnamefont {Haack}}, \bibinfo {author} {\bibfnamefont {J.~B.}\ \bibnamefont {Brask}}, \bibinfo {author} {\bibfnamefont {N.}~\bibnamefont {Brunner}}, \ and\ \bibinfo {author} {\bibfnamefont {M.}~\bibnamefont {Huber}},\ }\bibfield  {title} {\enquote {\bibinfo {title} {Unifying paradigms of quantum refrigeration: A universal and attainable bound on cooling},}\ }\href {\doibase 10.1103/PhysRevLett.123.170605} {\bibfield  {journal} {\bibinfo  {journal} {Phys. Rev. Lett.}\ }\textbf {\bibinfo {volume} {123}},\ \bibinfo {pages} {170605} (\bibinfo {year} {2019})}\BibitemShut {NoStop}%
\bibitem [{\citenamefont {Mitchison}(2019)}]{mitchison2019}%
  \BibitemOpen
  \bibfield  {author} {\bibinfo {author} {\bibfnamefont {M.~T.}\ \bibnamefont {Mitchison}},\ }\bibfield  {title} {\enquote {\bibinfo {title} {Quantum thermal absorption machines: Refrigerators, engines and clocks},}\ }\href {\doibase 10.1080/00107514.2019.1631555} {\bibfield  {journal} {\bibinfo  {journal} {Contemporary Physics}\ }\textbf {\bibinfo {volume} {60}},\ \bibinfo {pages} {164--187} (\bibinfo {year} {2019})}\BibitemShut {NoStop}%
\bibitem [{\citenamefont {Bhattacharyya}\ \emph {et~al.}(2025)\citenamefont {Bhattacharyya}, \citenamefont {Ghoshal},\ and\ \citenamefont {Sen}}]{apa_ref}%
  \BibitemOpen
  \bibfield  {author} {\bibinfo {author} {\bibfnamefont {A.}~\bibnamefont {Bhattacharyya}}, \bibinfo {author} {\bibfnamefont {A.}~\bibnamefont {Ghoshal}}, \ and\ \bibinfo {author} {\bibfnamefont {U.}~\bibnamefont {Sen}},\ }\bibfield  {title} {\enquote {\bibinfo {title} {Transient effects in quantum refrigerators with finite environments},}\ }\href {\doibase 10.1103/PhysRevA.111.012209} {\bibfield  {journal} {\bibinfo  {journal} {Phys. Rev. A}\ }\textbf {\bibinfo {volume} {111}},\ \bibinfo {pages} {012209} (\bibinfo {year} {2025})}\BibitemShut {NoStop}%
\bibitem [{\citenamefont {Hovhannisyan}\ \emph {et~al.}(2013{\natexlab{a}})\citenamefont {Hovhannisyan}, \citenamefont {Perarnau-Llobet}, \citenamefont {Huber},\ and\ \citenamefont {Ac\'{\i}n}}]{Acin}%
  \BibitemOpen
  \bibfield  {author} {\bibinfo {author} {\bibfnamefont {K.~V.}\ \bibnamefont {Hovhannisyan}}, \bibinfo {author} {\bibfnamefont {M.}~\bibnamefont {Perarnau-Llobet}}, \bibinfo {author} {\bibfnamefont {M.}~\bibnamefont {Huber}}, \ and\ \bibinfo {author} {\bibfnamefont {A.}~\bibnamefont {Ac\'{\i}n}},\ }\bibfield  {title} {\enquote {\bibinfo {title} {Entanglement generation is not necessary for optimal work extraction},}\ }\href {\doibase 10.1103/PhysRevLett.111.240401} {\bibfield  {journal} {\bibinfo  {journal} {Phys. Rev. Lett.}\ }\textbf {\bibinfo {volume} {111}},\ \bibinfo {pages} {240401} (\bibinfo {year} {2013}{\natexlab{a}})}\BibitemShut {NoStop}%
\bibitem [{\citenamefont {Frey}\ \emph {et~al.}(2014)\citenamefont {Frey}, \citenamefont {Funo},\ and\ \citenamefont {Hotta}}]{Frey}%
  \BibitemOpen
  \bibfield  {author} {\bibinfo {author} {\bibfnamefont {M.}~\bibnamefont {Frey}}, \bibinfo {author} {\bibfnamefont {K.}~\bibnamefont {Funo}}, \ and\ \bibinfo {author} {\bibfnamefont {M.}~\bibnamefont {Hotta}},\ }\bibfield  {title} {\enquote {\bibinfo {title} {Strong local passivity in finite quantum systems},}\ }\href {\doibase 10.1103/PhysRevE.90.012127} {\bibfield  {journal} {\bibinfo  {journal} {Phys. Rev. E}\ }\textbf {\bibinfo {volume} {90}},\ \bibinfo {pages} {012127} (\bibinfo {year} {2014})}\BibitemShut {NoStop}%
\bibitem [{\citenamefont {Campaioli}\ \emph {et~al.}(2017)\citenamefont {Campaioli}, \citenamefont {Pollock}, \citenamefont {Binder}, \citenamefont {C\'eleri}, \citenamefont {Goold}, \citenamefont {Vinjanampathy},\ and\ \citenamefont {Modi}}]{PhysRevLett.118.150601}%
  \BibitemOpen
  \bibfield  {author} {\bibinfo {author} {\bibfnamefont {F.}~\bibnamefont {Campaioli}}, \bibinfo {author} {\bibfnamefont {F.~A.}\ \bibnamefont {Pollock}}, \bibinfo {author} {\bibfnamefont {F.~C.}\ \bibnamefont {Binder}}, \bibinfo {author} {\bibfnamefont {L.}~\bibnamefont {C\'eleri}}, \bibinfo {author} {\bibfnamefont {J.}~\bibnamefont {Goold}}, \bibinfo {author} {\bibfnamefont {S.}~\bibnamefont {Vinjanampathy}}, \ and\ \bibinfo {author} {\bibfnamefont {K.}~\bibnamefont {Modi}},\ }\bibfield  {title} {\enquote {\bibinfo {title} {Enhancing the charging power of quantum batteries},}\ }\href {\doibase 10.1103/PhysRevLett.118.150601} {\bibfield  {journal} {\bibinfo  {journal} {Phys. Rev. Lett.}\ }\textbf {\bibinfo {volume} {118}},\ \bibinfo {pages} {150601} (\bibinfo {year} {2017})}\BibitemShut {NoStop}%
\bibitem [{\citenamefont {Ferraro}\ \emph {et~al.}(2018)\citenamefont {Ferraro}, \citenamefont {Campisi}, \citenamefont {Andolina}, \citenamefont {Pellegrini},\ and\ \citenamefont {Polini}}]{PhysRevLett.120.117702}%
  \BibitemOpen
  \bibfield  {author} {\bibinfo {author} {\bibfnamefont {D.}~\bibnamefont {Ferraro}}, \bibinfo {author} {\bibfnamefont {M.}~\bibnamefont {Campisi}}, \bibinfo {author} {\bibfnamefont {G.~M.}\ \bibnamefont {Andolina}}, \bibinfo {author} {\bibfnamefont {V.}~\bibnamefont {Pellegrini}}, \ and\ \bibinfo {author} {\bibfnamefont {M.}~\bibnamefont {Polini}},\ }\bibfield  {title} {\enquote {\bibinfo {title} {High-power collective charging of a solid-state quantum battery},}\ }\href {\doibase 10.1103/PhysRevLett.120.117702} {\bibfield  {journal} {\bibinfo  {journal} {Phys. Rev. Lett.}\ }\textbf {\bibinfo {volume} {120}},\ \bibinfo {pages} {117702} (\bibinfo {year} {2018})}\BibitemShut {NoStop}%
\bibitem [{\citenamefont {Andolina}\ \emph {et~al.}(2018)\citenamefont {Andolina}, \citenamefont {Farina}, \citenamefont {Mari}, \citenamefont {Pellegrini}, \citenamefont {Giovannetti},\ and\ \citenamefont {Polini}}]{PhysRevB.98.205423}%
  \BibitemOpen
  \bibfield  {author} {\bibinfo {author} {\bibfnamefont {G.~M.}\ \bibnamefont {Andolina}}, \bibinfo {author} {\bibfnamefont {D.}~\bibnamefont {Farina}}, \bibinfo {author} {\bibfnamefont {A.}~\bibnamefont {Mari}}, \bibinfo {author} {\bibfnamefont {V.}~\bibnamefont {Pellegrini}}, \bibinfo {author} {\bibfnamefont {V.}~\bibnamefont {Giovannetti}}, \ and\ \bibinfo {author} {\bibfnamefont {M.}~\bibnamefont {Polini}},\ }\bibfield  {title} {\enquote {\bibinfo {title} {Charger-mediated energy transfer in exactly solvable models for quantum batteries},}\ }\href {\doibase 10.1103/PhysRevB.98.205423} {\bibfield  {journal} {\bibinfo  {journal} {Phys. Rev. B}\ }\textbf {\bibinfo {volume} {98}},\ \bibinfo {pages} {205423} (\bibinfo {year} {2018})}\BibitemShut {NoStop}%
\bibitem [{\citenamefont {Alhambra}\ \emph {et~al.}(2019)\citenamefont {Alhambra}, \citenamefont {Styliaris}, \citenamefont {Rodr\'{\i}guez-Briones}, \citenamefont {Sikora},\ and\ \citenamefont {Mart\'{\i}n-Mart\'{\i}nez}}]{Alhambra}%
  \BibitemOpen
  \bibfield  {author} {\bibinfo {author} {\bibfnamefont {\'A.~M.}\ \bibnamefont {Alhambra}}, \bibinfo {author} {\bibfnamefont {G.}~\bibnamefont {Styliaris}}, \bibinfo {author} {\bibfnamefont {N.~A.}\ \bibnamefont {Rodr\'{\i}guez-Briones}}, \bibinfo {author} {\bibfnamefont {J.}~\bibnamefont {Sikora}}, \ and\ \bibinfo {author} {\bibfnamefont {E.}~\bibnamefont {Mart\'{\i}n-Mart\'{\i}nez}},\ }\bibfield  {title} {\enquote {\bibinfo {title} {Fundamental limitations to local energy extraction in quantum systems},}\ }\href {\doibase 10.1103/PhysRevLett.123.190601} {\bibfield  {journal} {\bibinfo  {journal} {Phys. Rev. Lett.}\ }\textbf {\bibinfo {volume} {123}},\ \bibinfo {pages} {190601} (\bibinfo {year} {2019})}\BibitemShut {NoStop}%
\bibitem [{\citenamefont {Santos}\ \emph {et~al.}(2019)\citenamefont {Santos}, \citenamefont {\ifmmode~\mbox{\c{C}}\else \c{C}\fi{}akmak}, \citenamefont {Campbell},\ and\ \citenamefont {Zinner}}]{Santos}%
  \BibitemOpen
  \bibfield  {author} {\bibinfo {author} {\bibfnamefont {A.~C.}\ \bibnamefont {Santos}}, \bibinfo {author} {\bibfnamefont {B.}~\bibnamefont {\ifmmode~\mbox{\c{C}}\else \c{C}\fi{}akmak}}, \bibinfo {author} {\bibfnamefont {S.}~\bibnamefont {Campbell}}, \ and\ \bibinfo {author} {\bibfnamefont {N.~T.}\ \bibnamefont {Zinner}},\ }\bibfield  {title} {\enquote {\bibinfo {title} {Stable adiabatic quantum batteries},}\ }\href {\doibase 10.1103/PhysRevE.100.032107} {\bibfield  {journal} {\bibinfo  {journal} {Phys. Rev. E}\ }\textbf {\bibinfo {volume} {100}},\ \bibinfo {pages} {032107} (\bibinfo {year} {2019})}\BibitemShut {NoStop}%
\bibitem [{\citenamefont {Dou}\ \emph {et~al.}(2020)\citenamefont {Dou}, \citenamefont {Wang},\ and\ \citenamefont {Sun}}]{Sun}%
  \BibitemOpen
  \bibfield  {author} {\bibinfo {author} {\bibfnamefont {F.~Q.}\ \bibnamefont {Dou}}, \bibinfo {author} {\bibfnamefont {Y.~J.}\ \bibnamefont {Wang}}, \ and\ \bibinfo {author} {\bibfnamefont {J.~A.}\ \bibnamefont {Sun}},\ }\bibfield  {title} {\enquote {\bibinfo {title} {Closed-loop three-level charged quantum battery},}\ }\href {\doibase 10.1209/0295-5075/131/43001} {\bibfield  {journal} {\bibinfo  {journal} {EPL (Europhysics Letters)}\ }\textbf {\bibinfo {volume} {131}},\ \bibinfo {pages} {43001} (\bibinfo {year} {2020})}\BibitemShut {NoStop}%
\bibitem [{\citenamefont {Ghosh}\ \emph {et~al.}(2020)\citenamefont {Ghosh}, \citenamefont {Chanda},\ and\ \citenamefont {Sen(De)}}]{Srijon1}%
  \BibitemOpen
  \bibfield  {author} {\bibinfo {author} {\bibfnamefont {S.}~\bibnamefont {Ghosh}}, \bibinfo {author} {\bibfnamefont {T.}~\bibnamefont {Chanda}}, \ and\ \bibinfo {author} {\bibfnamefont {A.}~\bibnamefont {Sen(De)}},\ }\bibfield  {title} {\enquote {\bibinfo {title} {Enhancement in the performance of a quantum battery by ordered and disordered interactions},}\ }\href {\doibase 10.1103/PhysRevA.101.032115} {\bibfield  {journal} {\bibinfo  {journal} {Phys. Rev. A}\ }\textbf {\bibinfo {volume} {101}},\ \bibinfo {pages} {032115} (\bibinfo {year} {2020})}\BibitemShut {NoStop}%
\bibitem [{\citenamefont {Rossini}\ \emph {et~al.}(2020)\citenamefont {Rossini}, \citenamefont {Andolina}, \citenamefont {Rosa}, \citenamefont {Carrega},\ and\ \citenamefont {Polini}}]{Polini0}%
  \BibitemOpen
  \bibfield  {author} {\bibinfo {author} {\bibfnamefont {D.}~\bibnamefont {Rossini}}, \bibinfo {author} {\bibfnamefont {G.~M.}\ \bibnamefont {Andolina}}, \bibinfo {author} {\bibfnamefont {D.}~\bibnamefont {Rosa}}, \bibinfo {author} {\bibfnamefont {M.}~\bibnamefont {Carrega}}, \ and\ \bibinfo {author} {\bibfnamefont {M.}~\bibnamefont {Polini}},\ }\bibfield  {title} {\enquote {\bibinfo {title} {Quantum advantage in the charging process of sachdev-ye-kitaev batteries},}\ }\href {\doibase 10.1103/PhysRevLett.125.236402} {\bibfield  {journal} {\bibinfo  {journal} {Phys. Rev. Lett.}\ }\textbf {\bibinfo {volume} {125}},\ \bibinfo {pages} {236402} (\bibinfo {year} {2020})}\BibitemShut {NoStop}%
\bibitem [{\citenamefont {Ghosh}\ \emph {et~al.}(2021)\citenamefont {Ghosh}, \citenamefont {Chanda}, \citenamefont {Mal},\ and\ \citenamefont {Sen(De)}}]{Srijon2}%
  \BibitemOpen
  \bibfield  {author} {\bibinfo {author} {\bibfnamefont {S.}~\bibnamefont {Ghosh}}, \bibinfo {author} {\bibfnamefont {T.}~\bibnamefont {Chanda}}, \bibinfo {author} {\bibfnamefont {S.}~\bibnamefont {Mal}}, \ and\ \bibinfo {author} {\bibfnamefont {A.}~\bibnamefont {Sen(De)}},\ }\bibfield  {title} {\enquote {\bibinfo {title} {Fast charging of a quantum battery assisted by noise},}\ }\href {\doibase 10.1103/PhysRevA.104.032207} {\bibfield  {journal} {\bibinfo  {journal} {Phys. Rev. A}\ }\textbf {\bibinfo {volume} {104}},\ \bibinfo {pages} {032207} (\bibinfo {year} {2021})}\BibitemShut {NoStop}%
\bibitem [{\citenamefont {Sen}\ and\ \citenamefont {Sen}(2021)}]{Kornikar}%
  \BibitemOpen
  \bibfield  {author} {\bibinfo {author} {\bibfnamefont {K.}~\bibnamefont {Sen}}\ and\ \bibinfo {author} {\bibfnamefont {U.}~\bibnamefont {Sen}},\ }\bibfield  {title} {\enquote {\bibinfo {title} {Local passivity and entanglement in shared quantum batteries},}\ }\href {\doibase 10.1103/PhysRevA.104.L030402} {\bibfield  {journal} {\bibinfo  {journal} {Phys. Rev. A}\ }\textbf {\bibinfo {volume} {104}},\ \bibinfo {pages} {L030402} (\bibinfo {year} {2021})}\BibitemShut {NoStop}%
\bibitem [{\citenamefont {Konar}\ \emph {et~al.}(2022)\citenamefont {Konar}, \citenamefont {Lakkaraju}, \citenamefont {Ghosh},\ and\ \citenamefont {Sen(De)}}]{Tanoy2}%
  \BibitemOpen
  \bibfield  {author} {\bibinfo {author} {\bibfnamefont {T.~K.}\ \bibnamefont {Konar}}, \bibinfo {author} {\bibfnamefont {L.~G.~C.}\ \bibnamefont {Lakkaraju}}, \bibinfo {author} {\bibfnamefont {S.}~\bibnamefont {Ghosh}}, \ and\ \bibinfo {author} {\bibfnamefont {A.}~\bibnamefont {Sen(De)}},\ }\bibfield  {title} {\enquote {\bibinfo {title} {Quantum battery with ultracold atoms: Bosons versus fermions},}\ }\href {\doibase 10.1103/PhysRevA.106.022618} {\bibfield  {journal} {\bibinfo  {journal} {Phys. Rev. A}\ }\textbf {\bibinfo {volume} {106}},\ \bibinfo {pages} {022618} (\bibinfo {year} {2022})}\BibitemShut {NoStop}%
\bibitem [{\citenamefont {Gyhm}\ and\ \citenamefont {Fischer}(2024)}]{fischer}%
  \BibitemOpen
  \bibfield  {author} {\bibinfo {author} {\bibfnamefont {J.~Y.}\ \bibnamefont {Gyhm}}\ and\ \bibinfo {author} {\bibfnamefont {U.~R.}\ \bibnamefont {Fischer}},\ }\bibfield  {title} {\enquote {\bibinfo {title} {{Beneficial and detrimental entanglement for quantum battery charging}},}\ }\href {https://doi.org/10.1116/5.0184903} {\bibfield  {journal} {\bibinfo  {journal} {AVS Quantum Science}\ }\textbf {\bibinfo {volume} {6}},\ \bibinfo {pages} {012001} (\bibinfo {year} {2024})}\BibitemShut {NoStop}%
\bibitem [{\citenamefont {Allahverdyan}\ \emph {et~al.}(2004)\citenamefont {Allahverdyan}, \citenamefont {Balian},\ and\ \citenamefont {Nieuwenhuizen}}]{allahverdyan2004maximal}%
  \BibitemOpen
  \bibfield  {author} {\bibinfo {author} {\bibfnamefont {A.~E}\ \bibnamefont {Allahverdyan}}, \bibinfo {author} {\bibfnamefont {R.}~\bibnamefont {Balian}}, \ and\ \bibinfo {author} {\bibfnamefont {T.~M}\ \bibnamefont {Nieuwenhuizen}},\ }\bibfield  {title} {\enquote {\bibinfo {title} {Maximal work extraction from finite quantum systems},}\ }\href {\doibase 10.1209/epl/i2004-10101-2} {\bibfield  {journal} {\bibinfo  {journal} {Europhysics Letters}\ }\textbf {\bibinfo {volume} {67}},\ \bibinfo {pages} {565} (\bibinfo {year} {2004})}\BibitemShut {NoStop}%
\bibitem [{\citenamefont {Perarnau-Llobet}\ \emph {et~al.}(2015{\natexlab{a}})\citenamefont {Perarnau-Llobet}, \citenamefont {Hovhannisyan}, \citenamefont {Huber}, \citenamefont {Skrzypczyk}, \citenamefont {Brunner},\ and\ \citenamefont {Ac\'{\i}n}}]{perarnau2015extractable}%
  \BibitemOpen
  \bibfield  {author} {\bibinfo {author} {\bibfnamefont {M.}~\bibnamefont {Perarnau-Llobet}}, \bibinfo {author} {\bibfnamefont {K.~V.}\ \bibnamefont {Hovhannisyan}}, \bibinfo {author} {\bibfnamefont {M.}~\bibnamefont {Huber}}, \bibinfo {author} {\bibfnamefont {P.}~\bibnamefont {Skrzypczyk}}, \bibinfo {author} {\bibfnamefont {Ni.}\ \bibnamefont {Brunner}}, \ and\ \bibinfo {author} {\bibfnamefont {A.}~\bibnamefont {Ac\'{\i}n}},\ }\bibfield  {title} {\enquote {\bibinfo {title} {Extractable work from correlations},}\ }\href {\doibase 10.1103/PhysRevX.5.041011} {\bibfield  {journal} {\bibinfo  {journal} {Phys. Rev. X}\ }\textbf {\bibinfo {volume} {5}},\ \bibinfo {pages} {041011} (\bibinfo {year} {2015}{\natexlab{a}})}\BibitemShut {NoStop}%
\bibitem [{\citenamefont {Binder}\ \emph {et~al.}(2015)\citenamefont {Binder}, \citenamefont {Vinjanampathy}, \citenamefont {Modi},\ and\ \citenamefont {Goold}}]{binder2015quantacell}%
  \BibitemOpen
  \bibfield  {author} {\bibinfo {author} {\bibfnamefont {F.~C}\ \bibnamefont {Binder}}, \bibinfo {author} {\bibfnamefont {S.}~\bibnamefont {Vinjanampathy}}, \bibinfo {author} {\bibfnamefont {K.}~\bibnamefont {Modi}}, \ and\ \bibinfo {author} {\bibfnamefont {J.}~\bibnamefont {Goold}},\ }\bibfield  {title} {\enquote {\bibinfo {title} {Quantacell: powerful charging of quantum batteries},}\ }\href {\doibase 10.1088/1367-2630/17/7/075015} {\bibfield  {journal} {\bibinfo  {journal} {New Journal of Physics}\ }\textbf {\bibinfo {volume} {17}},\ \bibinfo {pages} {075015} (\bibinfo {year} {2015})}\BibitemShut {NoStop}%
\bibitem [{\citenamefont {Hovhannisyan}\ \emph {et~al.}(2013{\natexlab{b}})\citenamefont {Hovhannisyan}, \citenamefont {Perarnau-Llobet}, \citenamefont {Huber},\ and\ \citenamefont {Ac\'{\i}n}}]{PhysRevLett.111.240401}%
  \BibitemOpen
  \bibfield  {author} {\bibinfo {author} {\bibfnamefont {K.~V.}\ \bibnamefont {Hovhannisyan}}, \bibinfo {author} {\bibfnamefont {M.}~\bibnamefont {Perarnau-Llobet}}, \bibinfo {author} {\bibfnamefont {M.}~\bibnamefont {Huber}}, \ and\ \bibinfo {author} {\bibfnamefont {A.}~\bibnamefont {Ac\'{\i}n}},\ }\bibfield  {title} {\enquote {\bibinfo {title} {Entanglement generation is not necessary for optimal work extraction},}\ }\href {\doibase 10.1103/PhysRevLett.111.240401} {\bibfield  {journal} {\bibinfo  {journal} {Phys. Rev. Lett.}\ }\textbf {\bibinfo {volume} {111}},\ \bibinfo {pages} {240401} (\bibinfo {year} {2013}{\natexlab{b}})}\BibitemShut {NoStop}%
\bibitem [{\citenamefont {Le}\ \emph {et~al.}(2018)\citenamefont {Le}, \citenamefont {Levinsen}, \citenamefont {Modi}, \citenamefont {Parish},\ and\ \citenamefont {Pollock}}]{le2018spin}%
  \BibitemOpen
  \bibfield  {author} {\bibinfo {author} {\bibfnamefont {T.~P.}\ \bibnamefont {Le}}, \bibinfo {author} {\bibfnamefont {J.}~\bibnamefont {Levinsen}}, \bibinfo {author} {\bibfnamefont {K.}~\bibnamefont {Modi}}, \bibinfo {author} {\bibfnamefont {M.~M.}\ \bibnamefont {Parish}}, \ and\ \bibinfo {author} {\bibfnamefont {F.~A.}\ \bibnamefont {Pollock}},\ }\bibfield  {title} {\enquote {\bibinfo {title} {Spin-chain model of a many-body quantum battery},}\ }\href {\doibase 10.1103/PhysRevA.97.022106} {\bibfield  {journal} {\bibinfo  {journal} {Phys. Rev. A}\ }\textbf {\bibinfo {volume} {97}},\ \bibinfo {pages} {022106} (\bibinfo {year} {2018})}\BibitemShut {NoStop}%
\bibitem [{\citenamefont {Andolina}\ \emph {et~al.}(2019)\citenamefont {Andolina}, \citenamefont {Keck}, \citenamefont {Mari}, \citenamefont {Campisi}, \citenamefont {Giovannetti},\ and\ \citenamefont {Polini}}]{andolina2019extractable}%
  \BibitemOpen
  \bibfield  {author} {\bibinfo {author} {\bibfnamefont {G.~M.}\ \bibnamefont {Andolina}}, \bibinfo {author} {\bibfnamefont {M.}~\bibnamefont {Keck}}, \bibinfo {author} {\bibfnamefont {A.}~\bibnamefont {Mari}}, \bibinfo {author} {\bibfnamefont {M.}~\bibnamefont {Campisi}}, \bibinfo {author} {\bibfnamefont {V.}~\bibnamefont {Giovannetti}}, \ and\ \bibinfo {author} {\bibfnamefont {M.}~\bibnamefont {Polini}},\ }\bibfield  {title} {\enquote {\bibinfo {title} {Extractable work, the role of correlations, and asymptotic freedom in quantum batteries},}\ }\href {\doibase 10.1103/PhysRevLett.122.047702} {\bibfield  {journal} {\bibinfo  {journal} {Phys. Rev. Lett.}\ }\textbf {\bibinfo {volume} {122}},\ \bibinfo {pages} {047702} (\bibinfo {year} {2019})}\BibitemShut {NoStop}%
\bibitem [{\citenamefont {Bhattacharyya}\ \emph {et~al.}(2024)\citenamefont {Bhattacharyya}, \citenamefont {Sen},\ and\ \citenamefont {Sen}}]{ncp_Aparajita}%
  \BibitemOpen
  \bibfield  {author} {\bibinfo {author} {\bibfnamefont {A.}~\bibnamefont {Bhattacharyya}}, \bibinfo {author} {\bibfnamefont {K.}~\bibnamefont {Sen}}, \ and\ \bibinfo {author} {\bibfnamefont {U.}~\bibnamefont {Sen}},\ }\bibfield  {title} {\enquote {\bibinfo {title} {Noncompletely positive quantum maps enable efficient local energy extraction in batteries},}\ }\href {\doibase 10.1103/PhysRevLett.132.240401} {\bibfield  {journal} {\bibinfo  {journal} {Phys. Rev. Lett.}\ }\textbf {\bibinfo {volume} {132}},\ \bibinfo {pages} {240401} (\bibinfo {year} {2024})}\BibitemShut {NoStop}%
\bibitem [{\citenamefont {Alicki}\ and\ \citenamefont {Fannes}(2013)}]{alicki2013entanglement}%
  \BibitemOpen
  \bibfield  {author} {\bibinfo {author} {\bibfnamefont {R.}~\bibnamefont {Alicki}}\ and\ \bibinfo {author} {\bibfnamefont {M.}~\bibnamefont {Fannes}},\ }\bibfield  {title} {\enquote {\bibinfo {title} {Entanglement boost for extractable work from ensembles of quantum batteries},}\ }\href {\doibase 10.1103/PhysRevE.87.042123} {\bibfield  {journal} {\bibinfo  {journal} {Phys. Rev. E}\ }\textbf {\bibinfo {volume} {87}},\ \bibinfo {pages} {042123} (\bibinfo {year} {2013})}\BibitemShut {NoStop}%
\bibitem [{\citenamefont {Crescente}\ \emph {et~al.}(2020{\natexlab{a}})\citenamefont {Crescente}, \citenamefont {Carrega}, \citenamefont {Sassetti},\ and\ \citenamefont {Ferraro}}]{crescente2020charging}%
  \BibitemOpen
  \bibfield  {author} {\bibinfo {author} {\bibfnamefont {A}~\bibnamefont {Crescente}}, \bibinfo {author} {\bibfnamefont {M}~\bibnamefont {Carrega}}, \bibinfo {author} {\bibfnamefont {M}~\bibnamefont {Sassetti}}, \ and\ \bibinfo {author} {\bibfnamefont {D}~\bibnamefont {Ferraro}},\ }\bibfield  {title} {\enquote {\bibinfo {title} {Charging and energy fluctuations of a driven quantum battery},}\ }\href {\doibase 10.1088/1367-2630/ab91fc} {\bibfield  {journal} {\bibinfo  {journal} {New Journal of Physics}\ }\textbf {\bibinfo {volume} {22}},\ \bibinfo {pages} {063057} (\bibinfo {year} {2020}{\natexlab{a}})}\BibitemShut {NoStop}%
\bibitem [{\citenamefont {Santos}\ \emph {et~al.}(2020)\citenamefont {Santos}, \citenamefont {Saguia},\ and\ \citenamefont {Sarandy}}]{PhysRevE.101.062114}%
  \BibitemOpen
  \bibfield  {author} {\bibinfo {author} {\bibfnamefont {A.~C.}\ \bibnamefont {Santos}}, \bibinfo {author} {\bibfnamefont {A.}~\bibnamefont {Saguia}}, \ and\ \bibinfo {author} {\bibfnamefont {M.~S.}\ \bibnamefont {Sarandy}},\ }\bibfield  {title} {\enquote {\bibinfo {title} {Stable and charge-switchable quantum batteries},}\ }\href {\doibase 10.1103/PhysRevE.101.062114} {\bibfield  {journal} {\bibinfo  {journal} {Phys. Rev. E}\ }\textbf {\bibinfo {volume} {101}},\ \bibinfo {pages} {062114} (\bibinfo {year} {2020})}\BibitemShut {NoStop}%
\bibitem [{\citenamefont {Carrega}\ \emph {et~al.}(2020)\citenamefont {Carrega}, \citenamefont {Crescente}, \citenamefont {Ferraro},\ and\ \citenamefont {Sassetti}}]{carrega2020dissipative}%
  \BibitemOpen
  \bibfield  {author} {\bibinfo {author} {\bibfnamefont {M.}~\bibnamefont {Carrega}}, \bibinfo {author} {\bibfnamefont {A.}~\bibnamefont {Crescente}}, \bibinfo {author} {\bibfnamefont {D.}~\bibnamefont {Ferraro}}, \ and\ \bibinfo {author} {\bibfnamefont {M.}~\bibnamefont {Sassetti}},\ }\bibfield  {title} {\enquote {\bibinfo {title} {Dissipative dynamics of an open quantum battery},}\ }\href {\doibase 10.1088/1367-2630/abaa01} {\bibfield  {journal} {\bibinfo  {journal} {New Journal of Physics}\ }\textbf {\bibinfo {volume} {22}},\ \bibinfo {pages} {083085} (\bibinfo {year} {2020})}\BibitemShut {NoStop}%
\bibitem [{\citenamefont {Santos}(2021)}]{PhysRevE.103.042118}%
  \BibitemOpen
  \bibfield  {author} {\bibinfo {author} {\bibfnamefont {A.~C.}\ \bibnamefont {Santos}},\ }\bibfield  {title} {\enquote {\bibinfo {title} {Quantum advantage of two-level batteries in the self-discharging process},}\ }\href {\doibase 10.1103/PhysRevE.103.042118} {\bibfield  {journal} {\bibinfo  {journal} {Phys. Rev. E}\ }\textbf {\bibinfo {volume} {103}},\ \bibinfo {pages} {042118} (\bibinfo {year} {2021})}\BibitemShut {NoStop}%
\bibitem [{\citenamefont {Shaghaghi}\ \emph {et~al.}(2022)\citenamefont {Shaghaghi}, \citenamefont {Singh}, \citenamefont {Benenti},\ and\ \citenamefont {Rosa}}]{shaghaghi2022micromasers}%
  \BibitemOpen
  \bibfield  {author} {\bibinfo {author} {\bibfnamefont {V.}~\bibnamefont {Shaghaghi}}, \bibinfo {author} {\bibfnamefont {V.}~\bibnamefont {Singh}}, \bibinfo {author} {\bibfnamefont {G.}~\bibnamefont {Benenti}}, \ and\ \bibinfo {author} {\bibfnamefont {D.}~\bibnamefont {Rosa}},\ }\bibfield  {title} {\enquote {\bibinfo {title} {Micromasers as quantum batteries},}\ }\href {\doibase 10.1088/2058-9565/ac8829} {\bibfield  {journal} {\bibinfo  {journal} {Quantum Science and Technology}\ }\textbf {\bibinfo {volume} {7}},\ \bibinfo {pages} {04LT01} (\bibinfo {year} {2022})}\BibitemShut {NoStop}%
\bibitem [{\citenamefont {Gyhm}\ \emph {et~al.}(2022)\citenamefont {Gyhm}, \citenamefont {\ifmmode~\check{S}\else \v{S}\fi{}afr\'anek},\ and\ \citenamefont {Rosa}}]{PhysRevLett.128.140501}%
  \BibitemOpen
  \bibfield  {author} {\bibinfo {author} {\bibfnamefont {J.}~\bibnamefont {Gyhm}}, \bibinfo {author} {\bibfnamefont {D.}~\bibnamefont {\ifmmode~\check{S}\else \v{S}\fi{}afr\'anek}}, \ and\ \bibinfo {author} {\bibfnamefont {D.}~\bibnamefont {Rosa}},\ }\bibfield  {title} {\enquote {\bibinfo {title} {Quantum charging advantage cannot be extensive without global operations},}\ }\href {\doibase 10.1103/PhysRevLett.128.140501} {\bibfield  {journal} {\bibinfo  {journal} {Phys. Rev. Lett.}\ }\textbf {\bibinfo {volume} {128}},\ \bibinfo {pages} {140501} (\bibinfo {year} {2022})}\BibitemShut {NoStop}%
\bibitem [{\citenamefont {Chaki}\ \emph {et~al.}(2023)\citenamefont {Chaki}, \citenamefont {Bhattacharyya}, \citenamefont {Sen},\ and\ \citenamefont {Sen}}]{chaki2023}%
  \BibitemOpen
  \bibfield  {author} {\bibinfo {author} {\bibfnamefont {P.}~\bibnamefont {Chaki}}, \bibinfo {author} {\bibfnamefont {A.}~\bibnamefont {Bhattacharyya}}, \bibinfo {author} {\bibfnamefont {K.}~\bibnamefont {Sen}}, \ and\ \bibinfo {author} {\bibfnamefont {U.}~\bibnamefont {Sen}},\ }\bibfield  {title} {\enquote {\bibinfo {title} {Auxiliary-assisted stochastic energy extraction from quantum batteries},}\ }\href {https://arxiv.org/abs/2307.16856} {\bibfield  {journal} {\bibinfo  {journal} {arXiv:2307.16856}\ } (\bibinfo {year} {2023})}\BibitemShut {NoStop}%
\bibitem [{\citenamefont {Chaki}\ \emph {et~al.}(2024{\natexlab{a}})\citenamefont {Chaki}, \citenamefont {Bhattacharyya}, \citenamefont {Sen},\ and\ \citenamefont {Sen}}]{chaki2024}%
  \BibitemOpen
  \bibfield  {author} {\bibinfo {author} {\bibfnamefont {P.}~\bibnamefont {Chaki}}, \bibinfo {author} {\bibfnamefont {A.}~\bibnamefont {Bhattacharyya}}, \bibinfo {author} {\bibfnamefont {K.}~\bibnamefont {Sen}}, \ and\ \bibinfo {author} {\bibfnamefont {U.}~\bibnamefont {Sen}},\ }\bibfield  {title} {\enquote {\bibinfo {title} {Positive and non-positive measurements in energy extraction from quantum batteries},}\ }\href {https://arxiv.org/abs/2404.18745} {\bibfield  {journal} {\bibinfo  {journal} {arXiv:2404.18745}\ } (\bibinfo {year} {2024}{\natexlab{a}})}\BibitemShut {NoStop}%
\bibitem [{\citenamefont {Chaki}\ \emph {et~al.}(2024{\natexlab{b}})\citenamefont {Chaki}, \citenamefont {Bhattacharyya},\ and\ \citenamefont {Sen}}]{cata}%
  \BibitemOpen
  \bibfield  {author} {\bibinfo {author} {\bibfnamefont {P.}~\bibnamefont {Chaki}}, \bibinfo {author} {\bibfnamefont {A.}~\bibnamefont {Bhattacharyya}}, \ and\ \bibinfo {author} {\bibfnamefont {U.}~\bibnamefont {Sen}},\ }\bibfield  {title} {\enquote {\bibinfo {title} {Universal and complete extraction for energy-invariant catalysis in quantum batteries versus no uncorrelated state-invariant catalysis},}\ }\href {https://arxiv.org/abs/2409.14153} {\bibfield  {journal} {\bibinfo  {journal} {arXiv:2409.14153}\ } (\bibinfo {year} {2024}{\natexlab{b}})}\BibitemShut {NoStop}%
\bibitem [{\citenamefont {Farina}\ \emph {et~al.}(2019)\citenamefont {Farina}, \citenamefont {Andolina}, \citenamefont {Mari}, \citenamefont {Polini},\ and\ \citenamefont {Giovannetti}}]{andolina2019charger}%
  \BibitemOpen
  \bibfield  {author} {\bibinfo {author} {\bibfnamefont {D.}~\bibnamefont {Farina}}, \bibinfo {author} {\bibfnamefont {G.~M.}\ \bibnamefont {Andolina}}, \bibinfo {author} {\bibfnamefont {A.}~\bibnamefont {Mari}}, \bibinfo {author} {\bibfnamefont {M.}~\bibnamefont {Polini}}, \ and\ \bibinfo {author} {\bibfnamefont {V.}~\bibnamefont {Giovannetti}},\ }\bibfield  {title} {\enquote {\bibinfo {title} {Charger-mediated energy transfer for quantum batteries: An open-system approach},}\ }\href {\doibase 10.1103/PhysRevB.99.035421} {\bibfield  {journal} {\bibinfo  {journal} {Phys. Rev. B}\ }\textbf {\bibinfo {volume} {99}},\ \bibinfo {pages} {035421} (\bibinfo {year} {2019})}\BibitemShut {NoStop}%
\bibitem [{\citenamefont {Downing}\ and\ \citenamefont {Ukhtary}(2023)}]{downing2023quantum}%
  \BibitemOpen
  \bibfield  {author} {\bibinfo {author} {\bibfnamefont {C.~A.}\ \bibnamefont {Downing}}\ and\ \bibinfo {author} {\bibfnamefont {M.~S.}\ \bibnamefont {Ukhtary}},\ }\bibfield  {title} {\enquote {\bibinfo {title} {A quantum battery with quadratic driving},}\ }\href {\doibase 10.1038/s42005-023-01439-y} {\bibfield  {journal} {\bibinfo  {journal} {Communications Physics}\ }\textbf {\bibinfo {volume} {6}},\ \bibinfo {pages} {322} (\bibinfo {year} {2023})}\BibitemShut {NoStop}%
\bibitem [{\citenamefont {Felicetti}\ \emph {et~al.}(2018)\citenamefont {Felicetti}, \citenamefont {Rossatto}, \citenamefont {Rico}, \citenamefont {Solano},\ and\ \citenamefont {Forn-Diaz}}]{Felicetti_2018}%
  \BibitemOpen
  \bibfield  {author} {\bibinfo {author} {\bibfnamefont {S.}~\bibnamefont {Felicetti}}, \bibinfo {author} {\bibfnamefont {D.~Z.}\ \bibnamefont {Rossatto}}, \bibinfo {author} {\bibfnamefont {E.}~\bibnamefont {Rico}}, \bibinfo {author} {\bibfnamefont {E.}~\bibnamefont {Solano}}, \ and\ \bibinfo {author} {\bibfnamefont {P.}~\bibnamefont {Forn-Diaz}},\ }\bibfield  {title} {\enquote {\bibinfo {title} {Two-photon quantum rabi model with superconducting circuits},}\ }\href {\doibase 10.1103/PhysRevA.97.013851} {\bibfield  {journal} {\bibinfo  {journal} {Physical Review A}\ }\textbf {\bibinfo {volume} {97}},\ \bibinfo {pages} {013851} (\bibinfo {year} {2018})}\BibitemShut {NoStop}%
\bibitem [{\citenamefont {Felicetti}\ \emph {et~al.}(2015)\citenamefont {Felicetti}, \citenamefont {Padernales}, \citenamefont {Egusquiza}, \citenamefont {Romero}, \citenamefont {Lamata}, \citenamefont {Braak},\ and\ \citenamefont {Solano}}]{Felicetti_2015}%
  \BibitemOpen
  \bibfield  {author} {\bibinfo {author} {\bibfnamefont {S.}~\bibnamefont {Felicetti}}, \bibinfo {author} {\bibfnamefont {J.~S.}\ \bibnamefont {Padernales}}, \bibinfo {author} {\bibfnamefont {I.~L.}\ \bibnamefont {Egusquiza}}, \bibinfo {author} {\bibfnamefont {G.}~\bibnamefont {Romero}}, \bibinfo {author} {\bibfnamefont {L.}~\bibnamefont {Lamata}}, \bibinfo {author} {\bibfnamefont {D.}~\bibnamefont {Braak}}, \ and\ \bibinfo {author} {\bibfnamefont {E.}~\bibnamefont {Solano}},\ }\bibfield  {title} {\enquote {\bibinfo {title} {Spectral collapse via two-phonon interactions in trapped ions},}\ }\href {\doibase 10.1103/PhysRevA.92.033817} {\bibfield  {journal} {\bibinfo  {journal} {Physical Review A}\ }\textbf {\bibinfo {volume} {92}},\ \bibinfo {pages} {033817} (\bibinfo {year} {2015})}\BibitemShut {NoStop}%
\bibitem [{\citenamefont {Garbe}\ \emph {et~al.}(2017)\citenamefont {Garbe}, \citenamefont {Egusquiza}, \citenamefont {Solano}, \citenamefont {Ciuti}, \citenamefont {Coudreau}, \citenamefont {Milman},\ and\ \citenamefont {Felicetti}}]{PhysRevA.95.053854}%
  \BibitemOpen
  \bibfield  {author} {\bibinfo {author} {\bibfnamefont {L.}~\bibnamefont {Garbe}}, \bibinfo {author} {\bibfnamefont {I.~L.}\ \bibnamefont {Egusquiza}}, \bibinfo {author} {\bibfnamefont {E.}~\bibnamefont {Solano}}, \bibinfo {author} {\bibfnamefont {C.}~\bibnamefont {Ciuti}}, \bibinfo {author} {\bibfnamefont {T.}~\bibnamefont {Coudreau}}, \bibinfo {author} {\bibfnamefont {P.}~\bibnamefont {Milman}}, \ and\ \bibinfo {author} {\bibfnamefont {S.}~\bibnamefont {Felicetti}},\ }\bibfield  {title} {\enquote {\bibinfo {title} {Superradiant phase transition in the ultrastrong-coupling regime of the two-photon dicke model},}\ }\href {\doibase 10.1103/PhysRevA.95.053854} {\bibfield  {journal} {\bibinfo  {journal} {Phys. Rev. A}\ }\textbf {\bibinfo {volume} {95}},\ \bibinfo {pages} {053854} (\bibinfo {year} {2017})}\BibitemShut {NoStop}%
\bibitem [{\citenamefont {Garbe}\ \emph {et~al.}(2020)\citenamefont {Garbe}, \citenamefont {Wade}, \citenamefont {Minganti}, \citenamefont {Shammah}, \citenamefont {Felicetti},\ and\ \citenamefont {Nori}}]{Garbe2020}%
  \BibitemOpen
  \bibfield  {author} {\bibinfo {author} {\bibfnamefont {Louis}\ \bibnamefont {Garbe}}, \bibinfo {author} {\bibfnamefont {Peregrine}\ \bibnamefont {Wade}}, \bibinfo {author} {\bibfnamefont {Fabrizio}\ \bibnamefont {Minganti}}, \bibinfo {author} {\bibfnamefont {Nathan}\ \bibnamefont {Shammah}}, \bibinfo {author} {\bibfnamefont {Simone}\ \bibnamefont {Felicetti}}, \ and\ \bibinfo {author} {\bibfnamefont {Franco}\ \bibnamefont {Nori}},\ }\bibfield  {title} {\enquote {\bibinfo {title} {Dissipation-induced bistability in the two-photon dicke model},}\ }\href {\doibase 10.1038/s41598-020-69704-6} {\bibfield  {journal} {\bibinfo  {journal} {Scientific Reports}\ }\textbf {\bibinfo {volume} {10}},\ \bibinfo {pages} {13408} (\bibinfo {year} {2020})}\BibitemShut {NoStop}%
\bibitem [{\citenamefont {Delmonte}\ \emph {et~al.}(2021)\citenamefont {Delmonte}, \citenamefont {Crescente}, \citenamefont {Carrega}, \citenamefont {Ferraro},\ and\ \citenamefont {Sassetti}}]{delmonte2021characterization}%
  \BibitemOpen
  \bibfield  {author} {\bibinfo {author} {\bibfnamefont {A.}~\bibnamefont {Delmonte}}, \bibinfo {author} {\bibfnamefont {A.}~\bibnamefont {Crescente}}, \bibinfo {author} {\bibfnamefont {M.}~\bibnamefont {Carrega}}, \bibinfo {author} {\bibfnamefont {D.}~\bibnamefont {Ferraro}}, \ and\ \bibinfo {author} {\bibfnamefont {M.}~\bibnamefont {Sassetti}},\ }\bibfield  {title} {\enquote {\bibinfo {title} {Characterization of a two-photon quantum battery: Initial conditions, stability and work extraction},}\ }\href {\doibase 10.3390/e23050612} {\bibfield  {journal} {\bibinfo  {journal} {Entropy}\ } (\bibinfo {year} {2021}),\ 10.3390/e23050612}\BibitemShut {NoStop}%
\bibitem [{\citenamefont {Crescente}\ \emph {et~al.}(2020{\natexlab{b}})\citenamefont {Crescente}, \citenamefont {Carrega}, \citenamefont {Sassetti},\ and\ \citenamefont {Ferraro}}]{PhysRevB.102.245407}%
  \BibitemOpen
  \bibfield  {author} {\bibinfo {author} {\bibfnamefont {A.}~\bibnamefont {Crescente}}, \bibinfo {author} {\bibfnamefont {M.}~\bibnamefont {Carrega}}, \bibinfo {author} {\bibfnamefont {M.}~\bibnamefont {Sassetti}}, \ and\ \bibinfo {author} {\bibfnamefont {D.}~\bibnamefont {Ferraro}},\ }\bibfield  {title} {\enquote {\bibinfo {title} {Ultrafast charging in a two-photon dicke quantum battery},}\ }\href {\doibase 10.1103/PhysRevB.102.245407} {\bibfield  {journal} {\bibinfo  {journal} {Phys. Rev. B}\ }\textbf {\bibinfo {volume} {102}},\ \bibinfo {pages} {245407} (\bibinfo {year} {2020}{\natexlab{b}})}\BibitemShut {NoStop}%
\bibitem [{\citenamefont {Breuer}\ and\ \citenamefont {Petruccione}(2002)}]{BRE02}%
  \BibitemOpen
  \bibfield  {author} {\bibinfo {author} {\bibfnamefont {H.~P.}\ \bibnamefont {Breuer}}\ and\ \bibinfo {author} {\bibfnamefont {F.}~\bibnamefont {Petruccione}},\ }\href@noop {} {\emph {\bibinfo {title} {The theory of open quantum systems}}}\ (\bibinfo  {publisher} {Oxford University Press},\ \bibinfo {address} {Great Clarendon Street},\ \bibinfo {year} {2002})\BibitemShut {NoStop}%
\bibitem [{\citenamefont {Rivas}\ and\ \citenamefont {Huelga}(2011)}]{rivasarxivcitation}%
  \BibitemOpen
  \bibfield  {author} {\bibinfo {author} {\bibfnamefont {A.}~\bibnamefont {Rivas}}\ and\ \bibinfo {author} {\bibfnamefont {S.~F.}\ \bibnamefont {Huelga}},\ }\bibfield  {title} {\enquote {\bibinfo {title} {Open quantum systems. an introduction},}\ }\href {https://arxiv.org/abs/1104.5242} {\bibfield  {journal} {\bibinfo  {journal} {arXiv:1104.5242}\ } (\bibinfo {year} {2011})}\BibitemShut {NoStop}%
\bibitem [{\citenamefont {Martinis}\ \emph {et~al.}(2002)\citenamefont {Martinis}, \citenamefont {Nam}, \citenamefont {Aumentado},\ and\ \citenamefont {Urbina}}]{PhysRevLett.89.117901}%
  \BibitemOpen
  \bibfield  {author} {\bibinfo {author} {\bibfnamefont {J.~M.}\ \bibnamefont {Martinis}}, \bibinfo {author} {\bibfnamefont {S.}~\bibnamefont {Nam}}, \bibinfo {author} {\bibfnamefont {J.}~\bibnamefont {Aumentado}}, \ and\ \bibinfo {author} {\bibfnamefont {C.}~\bibnamefont {Urbina}},\ }\bibfield  {title} {\enquote {\bibinfo {title} {Rabi oscillations in a large josephson-junction qubit},}\ }\href {\doibase 10.1103/PhysRevLett.89.117901} {\bibfield  {journal} {\bibinfo  {journal} {Phys. Rev. Lett.}\ }\textbf {\bibinfo {volume} {89}},\ \bibinfo {pages} {117901} (\bibinfo {year} {2002})}\BibitemShut {NoStop}%
\bibitem [{\citenamefont {Martinis}\ \emph {et~al.}(2005)\citenamefont {Martinis}, \citenamefont {Cooper}, \citenamefont {McDermott}, \citenamefont {Steffen}, \citenamefont {Ansmann}, \citenamefont {Osborn}, \citenamefont {Cicak}, \citenamefont {Oh}, \citenamefont {Pappas}, \citenamefont {Simmonds},\ and\ \citenamefont {Yu}}]{PhysRevLett.95.210503}%
  \BibitemOpen
  \bibfield  {author} {\bibinfo {author} {\bibfnamefont {John~M.}\ \bibnamefont {Martinis}}, \bibinfo {author} {\bibfnamefont {K.~B.}\ \bibnamefont {Cooper}}, \bibinfo {author} {\bibfnamefont {R.}~\bibnamefont {McDermott}}, \bibinfo {author} {\bibfnamefont {Matthias}\ \bibnamefont {Steffen}}, \bibinfo {author} {\bibfnamefont {Markus}\ \bibnamefont {Ansmann}}, \bibinfo {author} {\bibfnamefont {K.~D.}\ \bibnamefont {Osborn}}, \bibinfo {author} {\bibfnamefont {K.}~\bibnamefont {Cicak}}, \bibinfo {author} {\bibfnamefont {Seongshik}\ \bibnamefont {Oh}}, \bibinfo {author} {\bibfnamefont {D.~P.}\ \bibnamefont {Pappas}}, \bibinfo {author} {\bibfnamefont {R.~W.}\ \bibnamefont {Simmonds}}, \ and\ \bibinfo {author} {\bibfnamefont {Clare~C.}\ \bibnamefont {Yu}},\ }\bibfield  {title} {\enquote {\bibinfo {title} {Decoherence in josephson qubits from dielectric loss},}\ }\href {\doibase 10.1103/PhysRevLett.95.210503} {\bibfield  {journal} {\bibinfo  {journal} {Phys. Rev. Lett.}\ }\textbf {\bibinfo {volume} {95}},\ \bibinfo
  {pages} {210503} (\bibinfo {year} {2005})}\BibitemShut {NoStop}%
\bibitem [{\citenamefont {Wendin}(2017)}]{wendin2017quantum}%
  \BibitemOpen
  \bibfield  {author} {\bibinfo {author} {\bibfnamefont {G.}~\bibnamefont {Wendin}},\ }\bibfield  {title} {\enquote {\bibinfo {title} {Quantum information processing with superconducting circuits: a review},}\ }\href {\doibase 10.1088/1361-6633/aa7e1a} {\bibfield  {journal} {\bibinfo  {journal} {Reports on Progress in Physics}\ }\textbf {\bibinfo {volume} {80}},\ \bibinfo {pages} {106001} (\bibinfo {year} {2017})}\BibitemShut {NoStop}%
\bibitem [{\citenamefont {Skrzypczyk}\ \emph {et~al.}(2015)\citenamefont {Skrzypczyk}, \citenamefont {Silva},\ and\ \citenamefont {Brunner}}]{Silva}%
  \BibitemOpen
  \bibfield  {author} {\bibinfo {author} {\bibfnamefont {P.}~\bibnamefont {Skrzypczyk}}, \bibinfo {author} {\bibfnamefont {R.}~\bibnamefont {Silva}}, \ and\ \bibinfo {author} {\bibfnamefont {N.}~\bibnamefont {Brunner}},\ }\bibfield  {title} {\enquote {\bibinfo {title} {Passivity, complete passivity, and virtual temperatures},}\ }\href {\doibase 10.1103/PhysRevE.91.052133} {\bibfield  {journal} {\bibinfo  {journal} {Phys. Rev. E}\ }\textbf {\bibinfo {volume} {91}},\ \bibinfo {pages} {052133} (\bibinfo {year} {2015})}\BibitemShut {NoStop}%
\bibitem [{\citenamefont {Perarnau-Llobet}\ \emph {et~al.}(2015{\natexlab{b}})\citenamefont {Perarnau-Llobet}, \citenamefont {Hovhannisyan}, \citenamefont {Huber}, \citenamefont {Skrzypczyk}, \citenamefont {Tura},\ and\ \citenamefont {Ac\'{\i}n}}]{Huber}%
  \BibitemOpen
  \bibfield  {author} {\bibinfo {author} {\bibfnamefont {M.}~\bibnamefont {Perarnau-Llobet}}, \bibinfo {author} {\bibfnamefont {K.~V.}\ \bibnamefont {Hovhannisyan}}, \bibinfo {author} {\bibfnamefont {M.}~\bibnamefont {Huber}}, \bibinfo {author} {\bibfnamefont {P.}~\bibnamefont {Skrzypczyk}}, \bibinfo {author} {\bibfnamefont {J.}~\bibnamefont {Tura}}, \ and\ \bibinfo {author} {\bibfnamefont {A.}~\bibnamefont {Ac\'{\i}n}},\ }\bibfield  {title} {\enquote {\bibinfo {title} {Most energetic passive states},}\ }\href {\doibase 10.1103/PhysRevE.92.042147} {\bibfield  {journal} {\bibinfo  {journal} {Phys. Rev. E}\ }\textbf {\bibinfo {volume} {92}},\ \bibinfo {pages} {042147} (\bibinfo {year} {2015}{\natexlab{b}})}\BibitemShut {NoStop}%
\bibitem [{\citenamefont {Brown}\ \emph {et~al.}(2016)\citenamefont {Brown}, \citenamefont {Friis},\ and\ \citenamefont {Huber}}]{Brown_2016}%
  \BibitemOpen
  \bibfield  {author} {\bibinfo {author} {\bibfnamefont {E.~G.}\ \bibnamefont {Brown}}, \bibinfo {author} {\bibfnamefont {N.}~\bibnamefont {Friis}}, \ and\ \bibinfo {author} {\bibfnamefont {M.}~\bibnamefont {Huber}},\ }\bibfield  {title} {\enquote {\bibinfo {title} {Passivity and practical work extraction using gaussian operations},}\ }\href {\doibase 10.1088/1367-2630/18/11/113028} {\bibfield  {journal} {\bibinfo  {journal} {New Journal of Physics}\ }\textbf {\bibinfo {volume} {18}},\ \bibinfo {pages} {113028} (\bibinfo {year} {2016})}\BibitemShut {NoStop}%
\bibitem [{\citenamefont {Sparaciari}\ \emph {et~al.}(2017)\citenamefont {Sparaciari}, \citenamefont {Jennings},\ and\ \citenamefont {Oppenheim}}]{Sparaciari_2017}%
  \BibitemOpen
  \bibfield  {author} {\bibinfo {author} {\bibfnamefont {C.}~\bibnamefont {Sparaciari}}, \bibinfo {author} {\bibfnamefont {D.}~\bibnamefont {Jennings}}, \ and\ \bibinfo {author} {\bibfnamefont {J.}~\bibnamefont {Oppenheim}},\ }\bibfield  {title} {\enquote {\bibinfo {title} {Energetic instability of passive states in thermodynamics},}\ }\href {https://doi.org/10.1038%2Fs41467-017-01505-4} {\bibfield  {journal} {\bibinfo  {journal} {Nature Communications}\ }\textbf {\bibinfo {volume} {8}} (\bibinfo {year} {2017})}\BibitemShut {NoStop}%
\bibitem [{\citenamefont {Onuma-Kalu}\ and\ \citenamefont {Mann}(2018)}]{Kalu}%
  \BibitemOpen
  \bibfield  {author} {\bibinfo {author} {\bibfnamefont {M.}~\bibnamefont {Onuma-Kalu}}\ and\ \bibinfo {author} {\bibfnamefont {R.~B.}\ \bibnamefont {Mann}},\ }\bibfield  {title} {\enquote {\bibinfo {title} {Work extraction using gaussian operations in noninteracting fermionic systems},}\ }\href {\doibase 10.1103/PhysRevE.98.042121} {\bibfield  {journal} {\bibinfo  {journal} {Phys. Rev. E}\ }\textbf {\bibinfo {volume} {98}},\ \bibinfo {pages} {042121} (\bibinfo {year} {2018})}\BibitemShut {NoStop}%
\bibitem [{\citenamefont {{Lenard}}(1978)}]{Lenard}%
  \BibitemOpen
  \bibfield  {author} {\bibinfo {author} {\bibfnamefont {A.}~\bibnamefont {{Lenard}}},\ }\bibfield  {title} {\enquote {\bibinfo {title} {{Thermodynamical proof of the Gibbs formula for elementary quantum systems}},}\ }\href {\doibase 10.1007/BF01011769} {\bibfield  {journal} {\bibinfo  {journal} {Journal of Statistical Physics}\ }\textbf {\bibinfo {volume} {19}},\ \bibinfo {pages} {575} (\bibinfo {year} {1978})}\BibitemShut {NoStop}%
\bibitem [{\citenamefont {Pusz}\ and\ \citenamefont {Woronowicz}(1978)}]{Pusz}%
  \BibitemOpen
  \bibfield  {author} {\bibinfo {author} {\bibfnamefont {W.}~\bibnamefont {Pusz}}\ and\ \bibinfo {author} {\bibfnamefont {S.~L.}\ \bibnamefont {Woronowicz}},\ }\bibfield  {title} {\enquote {\bibinfo {title} {{Passive states and KMS states for general quantum systems}},}\ }\href {https://link.springer.com/article/10.1007/bf01614224#citeas} {\bibfield  {journal} {\bibinfo  {journal} {Communications in Mathematical Physics}\ }\textbf {\bibinfo {volume} {58}},\ \bibinfo {pages} {273} (\bibinfo {year} {1978})}\BibitemShut {NoStop}%
\bibitem [{\citenamefont {Alicki}\ and\ \citenamefont {Lendi}(2007)}]{AlickiLendi2007}%
  \BibitemOpen
  \bibfield  {author} {\bibinfo {author} {\bibfnamefont {R.}~\bibnamefont {Alicki}}\ and\ \bibinfo {author} {\bibfnamefont {K.}~\bibnamefont {Lendi}},\ }\href@noop {} {\emph {\bibinfo {title} {Quantum Dynamical Semigroups and Applications}}}\ (\bibinfo  {publisher} {Springer},\ \bibinfo {address} {Berlin Heidelberg},\ \bibinfo {year} {2007})\BibitemShut {NoStop}%
\bibitem [{\citenamefont {Lidar}(2020)}]{lidar2020lecture}%
  \BibitemOpen
  \bibfield  {author} {\bibinfo {author} {\bibfnamefont {D.~A.}\ \bibnamefont {Lidar}},\ }\bibfield  {title} {\enquote {\bibinfo {title} {Lecture notes on the theory of open quantum systems},}\ }\href {https://arxiv.org/abs/1902.00967} {\bibfield  {journal} {\bibinfo  {journal} {arXiv:1902.00967}\ } (\bibinfo {year} {2020})}\BibitemShut {NoStop}%
\bibitem [{\citenamefont {Kossakowski}(1972)}]{Kossakowski1972}%
  \BibitemOpen
  \bibfield  {author} {\bibinfo {author} {\bibfnamefont {A.}~\bibnamefont {Kossakowski}},\ }\bibfield  {title} {\enquote {\bibinfo {title} {On quantum statistical mechanics of non-hamiltonian systems},}\ }\href {\doibase 10.1016/0034-4877(72)90010-9} {\bibfield  {journal} {\bibinfo  {journal} {Reports on Mathematical Physics}\ }\textbf {\bibinfo {volume} {3}},\ \bibinfo {pages} {247--274} (\bibinfo {year} {1972})}\BibitemShut {NoStop}%
\bibitem [{\citenamefont {Gorini}\ \emph {et~al.}(1976)\citenamefont {Gorini}, \citenamefont {Kossakowski},\ and\ \citenamefont {Sudarshan}}]{10.1063/1.522979}%
  \BibitemOpen
  \bibfield  {author} {\bibinfo {author} {\bibfnamefont {V.}~\bibnamefont {Gorini}}, \bibinfo {author} {\bibfnamefont {A.}~\bibnamefont {Kossakowski}}, \ and\ \bibinfo {author} {\bibfnamefont {E.~C.~G.}\ \bibnamefont {Sudarshan}},\ }\bibfield  {title} {\enquote {\bibinfo {title} {Completely positive dynamical semigroups of n‐level systems},}\ }\href {\doibase 10.1063/1.522979} {\bibfield  {journal} {\bibinfo  {journal} {Journal of Mathematical Physics}\ }\textbf {\bibinfo {volume} {17}},\ \bibinfo {pages} {821--825} (\bibinfo {year} {1976})}\BibitemShut {NoStop}%
\bibitem [{\citenamefont {Lindblad}(1976)}]{Lindblad1976}%
  \BibitemOpen
  \bibfield  {author} {\bibinfo {author} {\bibfnamefont {G.}~\bibnamefont {Lindblad}},\ }\bibfield  {title} {\enquote {\bibinfo {title} {On the generators of quantum dynamical semigroups},}\ }\href {\doibase 10.1007/BF01608499} {\bibfield  {journal} {\bibinfo  {journal} {Communications in Mathematical Physics}\ }\textbf {\bibinfo {volume} {48}},\ \bibinfo {pages} {119--130} (\bibinfo {year} {1976})}\BibitemShut {NoStop}%
\bibitem [{\citenamefont {Manzano}(2020)}]{Manzano_2020}%
  \BibitemOpen
  \bibfield  {author} {\bibinfo {author} {\bibfnamefont {D.}~\bibnamefont {Manzano}},\ }\bibfield  {title} {\enquote {\bibinfo {title} {A short introduction to the lindblad master equation},}\ }\href {\doibase 10.1063/1.5115323} {\bibfield  {journal} {\bibinfo  {journal} {AIP Advances}\ }\textbf {\bibinfo {volume} {10}} (\bibinfo {year} {2020}),\ 10.1063/1.5115323}\BibitemShut {NoStop}%
\bibitem [{\citenamefont {Nakamura}\ \emph {et~al.}(1999)\citenamefont {Nakamura}, \citenamefont {Pashkin},\ and\ \citenamefont {Tsai}}]{nakamura1999coherent}%
  \BibitemOpen
  \bibfield  {author} {\bibinfo {author} {\bibfnamefont {Y.}~\bibnamefont {Nakamura}}, \bibinfo {author} {\bibfnamefont {Y.}~\bibnamefont {Pashkin}}, \ and\ \bibinfo {author} {\bibfnamefont {J.}~\bibnamefont {Tsai}},\ }\bibfield  {title} {\enquote {\bibinfo {title} {Coherent control of macroscopic quantum states in a single-cooper-pair box},}\ }\href {\doibase 10.1038/19718} {\bibfield  {journal} {\bibinfo  {journal} {Nature}\ }\textbf {\bibinfo {volume} {398}},\ \bibinfo {pages} {786--788} (\bibinfo {year} {1999})}\BibitemShut {NoStop}%
\bibitem [{\citenamefont {Koch}\ \emph {et~al.}(2007)\citenamefont {Koch}, \citenamefont {Yu}, \citenamefont {Gambetta}, \citenamefont {Houck}, \citenamefont {Schuster}, \citenamefont {Majer}, \citenamefont {Blais}, \citenamefont {Devoret}, \citenamefont {Girvin},\ and\ \citenamefont {Schoelkopf}}]{PhysRevA.76.042319}%
  \BibitemOpen
  \bibfield  {author} {\bibinfo {author} {\bibfnamefont {J.}~\bibnamefont {Koch}}, \bibinfo {author} {\bibfnamefont {T.~M.}\ \bibnamefont {Yu}}, \bibinfo {author} {\bibfnamefont {J.}~\bibnamefont {Gambetta}}, \bibinfo {author} {\bibfnamefont {A.~A.}\ \bibnamefont {Houck}}, \bibinfo {author} {\bibfnamefont {D.~I.}\ \bibnamefont {Schuster}}, \bibinfo {author} {\bibfnamefont {J.}~\bibnamefont {Majer}}, \bibinfo {author} {\bibfnamefont {A.}~\bibnamefont {Blais}}, \bibinfo {author} {\bibfnamefont {M.~H.}\ \bibnamefont {Devoret}}, \bibinfo {author} {\bibfnamefont {S.~M.}\ \bibnamefont {Girvin}}, \ and\ \bibinfo {author} {\bibfnamefont {R.~J.}\ \bibnamefont {Schoelkopf}},\ }\bibfield  {title} {\enquote {\bibinfo {title} {Charge-insensitive qubit design derived from the cooper pair box},}\ }\href {\doibase 10.1103/PhysRevA.76.042319} {\bibfield  {journal} {\bibinfo  {journal} {Phys. Rev. A}\ }\textbf {\bibinfo {volume} {76}},\ \bibinfo {pages} {042319} (\bibinfo {year} {2007})}\BibitemShut {NoStop}%
\bibitem [{\citenamefont {Devoret}(1997)}]{devoret1997quantum}%
  \BibitemOpen
  \bibfield  {author} {\bibinfo {author} {\bibfnamefont {M.~H.}\ \bibnamefont {Devoret}},\ }\bibfield  {title} {\enquote {\bibinfo {title} {Quantum fluctuations},}\ }in\ \href@noop {} {\emph {\bibinfo {booktitle} {Les Houches Session LXIII}}},\ \bibinfo {editor} {edited by\ \bibinfo {editor} {\bibfnamefont {S.}~\bibnamefont {Reynaud}}, \bibinfo {editor} {\bibfnamefont {E.}~\bibnamefont {Giacobino}}, \ and\ \bibinfo {editor} {\bibfnamefont {J.}~\bibnamefont {Zinn-Justin}}}\ (\bibinfo  {publisher} {Elsevier},\ \bibinfo {address} {New York},\ \bibinfo {year} {1997})\ pp.\ \bibinfo {pages} {351--386}\BibitemShut {NoStop}%
\bibitem [{\citenamefont {Vool}\ and\ \citenamefont {Devoret}()}]{vool2017introduction}%
  \BibitemOpen
  \bibfield  {author} {\bibinfo {author} {\bibfnamefont {U.}~\bibnamefont {Vool}}\ and\ \bibinfo {author} {\bibfnamefont {M.}~\bibnamefont {Devoret}},\ }\bibfield  {title} {\enquote {\bibinfo {title} {Introduction to quantum electromagnetic circuits},}\ }\href {\doibase 10.1002/cta.2359} {\ 10.1002/cta.2359}\BibitemShut {NoStop}%
\end{thebibliography}%
\end{document}